\RequirePackage{xcolor,ifpdf}
\documentclass[hyper,letterpaper]{JHEP3}
\pdfoutput=1
\usepackage{tikz}
\usepackage{graphicx}

\usepackage{amsmath,amssymb,amsfonts,bm,empheq}
\usepackage{cite}
\usepackage{subcaption}
\usepackage{nicefrac}
\usepackage[enableskew,vcentermath]{youngtab}
\usepackage{epstopdf}
\usepackage{url}
\usepackage{float}
\usepackage{slashed,framed,xcolor,tikz}

\newcommand{\be}{\begin{equation}}
\newcommand{\ee}{\end{equation}}

\newcommand{\HFK}{\widehat{\mathrm{HFK}}}

\title{Revisiting the Melvin-Morton-Rozansky Expansion, \\ or There and Back Again}

\author{Sibasish Banerjee$^{1}$, Jakub Jankowski$^{2}$ and Piotr Su{\l}kowski$^{2,3}$
\\
$^1$ Department of Mathematics, Universit{\"a}t zu K{\"o}ln, Weyertal 86-90, D-50931, \\ Cologne, Germany \\
$^2$ Faculty of Physics, University of Warsaw, ul. Pasteura 5, 02-093 Warsaw, Poland \\
$^3$ Walter Burke Institute for Theoretical Physics, California Institute of Technology, Pasadena, CA 91125, USA 
}

\abstract{Alexander polynomial arises in the leading term of a semi-classical Melvin-Morton-Rozansky expansion of colored knot polynomials. In this work, following the opposite direction, we propose how to reconstruct colored HOMFLY-PT polynomials, superpolynomials, and newly introduced $\widehat{Z}$ invariants for some knot complements, from an appropriate rewriting, quantization and deformation of Alexander polynomial. Along this route we rederive conjectural expressions for the above mentioned invariants for various knots obtained recently, thereby proving their consistency with the Melvin-Morton-Rozansky theorem, and derive new formulae for colored superpolynomials unknown before. For a given knot, depending on certain choices, our reconstruction leads to equivalent expressions, which are either cyclotomic, or encode certain features of HOMFLY-PT homology and the knots-quivers correspondence.
\\
\\
\\
\\
\\
\\
\\
\\
\\
\\
\\
\\
\\
\\
\\
CALT-2020-030}

\begin{document}

\tableofcontents

\newpage
\section{Introduction}

Alexander polynomial $\Delta(x)$, the first polynomial knot invariant introduced almost a hundred years ago, plays a prominent role in both older and most recent developments in knot theory and its physical incarnations. Its essential role is revealed, for example, by the Melvin-Morton-Rozansky conjecture, subsequently turned into a theorem, which states that quantum $SU(N)$ knot invariants $P_r(q^N,q)$ colored by symmetric representations $S^r$ have the following semi-classical expansion
\be 
  \lim_{\hbar\to 0, r\to\infty}  P_r(q^N,q=e^{\hbar})\simeq \frac{1}{\Delta(x)^{N-1}} + \sum_{k=1}^\infty \frac{R_k(x,N)}{\Delta^{N+2k-1}(x)} \hbar^k ,\qquad\quad \textrm{for } x=q^r=const,
    \label{mmr-intro}
\ee
where $R_k(x,N)$ are polynomials in $x$. This conjecture was originally formulated by Melvin and Morton in the case of colored Jones polynomial (i.e. for $N=2$) \cite{Melvin}, and later generalized by Rozansky to more general knot polynomials \cite{Rozansky,Rozansky1,Rozansky2}. It has an important interpretation in physics, namely the above limit captures contributions to Chern-Simons partition function from an abelian flat connection. Its proof is given in \cite{Bar-Natan1996}, and an analytic version has been formulated and proven in \cite{garoufalidis2005analytic}. 

These days an interest in the Melvin-Morton-Rozansky expansion resurfaces. For example, it is believed that a general knot homology theory exists, which encompasses both knot Floer homology -- which categorifies Alexander polynomial -- and a putative (colored) HOMFLY-PT homology \cite{Dunfield:2005si,Gukov:2011ry}. In particular, structural properties of this latter homology theory recently enabled to derive conjectural formulae for colored HOMFLY-PT polynomials and superpolynomials \cite{Fuji:2012pm,Fuji:2012pi,Nawata:2012pg,Nawata:2015wya,Gukov:2015gmm}. The Melvin-Morton-Rozansky expansion, which relates $a=q^N$ specialization of colored HOMFLY-PT polynomials and Alexander polynomial, should therefore be one manifestation of such a general theory on a decategorified level. On the other hand, Melvin-Morton-Rozansky expansion enters a definition of newly introduced $\widehat{Z}$ invariants for knots complements, also denoted $F_K(x,q)$ or $F_K(x,a,q)$ \cite{Gukov:2019mnk,Park:2019xey,Ekholm:2020lqy}. These invariants are also inherently related to physics, as they count BPS states in a 3d $\mathcal{N}=2$ theory arising from compactification of 6d $(2,0)$ theory on a complement of a knot under consideration in $S^3$. It is thus worth revisiting the Melvin-Morton-Rozansky expansion and, among others, understanding its role in these new developments.

The main goal of this note is to understand in detail how the Melvin-Morton-Rozansky expansion arises from various recently found expressions for colored knot polynomials, and -- taking the opposite direction -- to provide prescription how to reconstruct colored HOMFLY-PT polynomials and superpolynomials starting from Alexander polynomial. We find that this reconstruction process can be split into two steps. First, we write the inverse of Alexander polynomial, i.e. the first term in the expansion (\ref{mmr-intro}), as a series of some particular form, essentially using the inverse binomial theorem. Second, we propose how to $q$-deform, $a$-deform, and possibly $t$-deform various terms in such a series. Depending on some choices made in the first step, the expressions that we obtain are either cyclotomic \cite{Habiro_2007,berest2019cyclotomic,Nawata:2015wya} -- which at the same time confirms the existence of cyclotomic expansions for colored HOMFLY-PT polynomials and superpolynomials -- or take equivalent form that encodes certain features of HOMFLY-PT homology and is relevant in the context of knots-quivers correspondence \cite{Kucharski:2017poe,Kucharski:2017ogk}. Making contact with cyclotomic expansions  and knots-quivers correspondence  is also an important aspect of this work. The second step, which captures $a$-, $q$- and $t$-deformation, involves a little arbitrariness, which can be fixed by comparison with known HOMFLY-PT polynomials for small colors, or (at least in principle) with first few polynomials $R_k(x,N)$ in the subleading terms in (\ref{mmr-intro}). Interestingly, to fix this arbitrariness in all examples that we consider in this paper, it is sufficient to compare just with polynomials colored by the fundamental and the second symmetric representation.

Note that some of our results reproduce conjectured expressions for colored knot polynomials found e.g. in \cite{Fuji:2012nx,Fuji:2012pi,Nawata:2012pg,Nawata:2015wya}. However, we stress that in those cases our derivation does not only provide an equivalent conjectural expressions, but also asserts that those expressions indeed satisfy the Melvin-Morton-Rozansky theorem. This is important because it is a new and independent check that these expressions are correct. Moreover, recall that those previously found expressions for colored polynomials  were derived taking advantage of the exponential growth condition \cite{Fuji:2012pm,Fuji:2012pi,Nawata:2012pg,Nawata:2015wya,Gukov:2015gmm}, i.e. the statement that colored superpolynomials $P_r(a,q,t)$ satisfy the relation $P_r(a,1,t)=P_{r=1}(a,1,t)^r$. Rewriting the right hand side of this relation in terms of binomial expansion leads to analogous expressions to those that we obtain, and ultimately the same results. However, the exponential growth property holds only for a restricted family of knots (in particular the thin knots), so it enables to determine colored polynomials only for this restricted class. On the other hand, the Melvin-Morton-Rozansky theorem that we take advantage of is proven for all knots, and one might expect that our techniques enable to get colored polynomials also for these knots that do not satisfy the exponential growth. 

To illustrate the power of our formalism, we also derive new formulae for colored polynomials, unknown before. In particular we focus on $7_4$ knot, which is one of the first knots for which explicit colored polynomials have not been found to date. Indeed, colored superpolynomials for knots up to 5 crossings were given in \cite{Fuji:2012pi}. Colored polynomials for twist knots, so in particular for $6_1$ and $7_2$, are found in \cite{Fuji:2012pi,Nawata:2012pg}. Formulae for $T^{2,2p+1}$ torus knots for all $p$, so in particular for $7_1$ knot, are given in \cite{Fuji:2012pi}. Colored superpolynomials for $6_2$ and $6_3$ knots are found in \cite{Nawata:2015wya}. Therefore the first knots for which colored polynomials have not been written down before are $7_3$, $7_4$, etc. We focus on derivation of colored superpolynomials for $7_4$ as an illustrative example, among others because it has the same Alexander polynomial as $9_2$ twist knot -- it is therefore instructive to see how our reconstruction process differs for those two knots. Following our prescription, one can analogously derive colored superpolynomials for other knots with 7 and more crossings.

In the course of our analysis we also find an interesting relation between colored HOMFLY-PT polynomials or superpolynomials for various pairs of knots, whose homological diagrams differ only by one element: a zig-zag of length 3 for one knot is replaced by a diamond and a trivial zig-zag of length 1 for the other knot. In consequence, Alexander polynomials for these knots, written as polynomials in $X=\frac{(1-x)^2}{x}$, differ only by signs. In turn, this leads to very similar cyclotomic expansions for such pairs of knots, which differ only by certain signs and deformation terms that arise in the second step of our reconstruction process, as mentioned above. The simplest pair of such knots are $3_1$ and $4_1$, whose Alexander polynomials can be written respectively as $\Delta_{3_1}(x)=1+X$ and $\Delta_{4_1}=1-X$. More generally, twist knots in the series $4_1,6_1,8_1,\ldots$ (labeled by $p=1,2,3,\ldots$), whose Alexander polynomials take form $\Delta(x)=1-pX$, are closely related to twist knots $3_1,5_2,7_2,\ldots$ (also labeled by $p=1,2,3,\ldots$), whose Alexander polynomials read $\Delta(x)=1+pX$. Similarly, Alexander polynomials for $6_2$ and $6_3$ knots differ only by signs, $\Delta_{6_2}(x)=1-X-X^2$ and $\Delta_{6_3}(x)=1+X+X^2$. It would be interesting to unveil other consequences of this correspondence.

Apart from reconstructing formulae for colored polynomials and understanding their structure, another important motivation for this work is that along similar lines expressions for $\widehat{Z}$ for knot complements might be found. For a complement of a knot $K$ in $S^3$, such invariants are also denoted $F_K(x,q)$, and they were originally introduced in \cite{Gukov:2019mnk} in the case corresponding to Jones polynomial. In this original setting, these invariants are supposed to have the same expansion (\ref{mmr-intro}) in the limit $\hbar\to 0$ (with $N=2$), and to satisfy the same quantum A-polynomial equation as colored Jones polynomials. A generalization of these invariants to $SU(N)$ case was proposed in \cite{Park:2019xey}, and $a$-deformed $F_K(x,a,q)$ invariants were introduced in \cite{Ekholm:2020lqy}. The latter invariants satisfy $a$-deformed quantum A-polynomial equations, the same as colored HOMFLY-PT polynomials \cite{Fuji:2012nx,Fuji:2013rra,Aganagic:2012jb}, and upon specialization $a=q^N$ have the same semi-classical limit as in (\ref{mmr-intro}). Therefore one might hope that $F_K$ invariants could be reconstructed following similar steps as we present in this work. We show that this is indeed the case at least for a family of $T^{2,2p+1}$ torus knots, for all $p$, as in this case $F_K(x,a,q)$ are closely related to colored polynomials (as also explained in \cite{Ekholm:2020lqy}). 

Finally, let us mention a few other problems worth pursuing, related to the results of this paper. The first interesting issue is an interplay between expansions in $\hbar$ or $q=e^{\hbar}$. In this work we postulate how to reconstruct colored polynomials or $F_K(x,q)$ invariants, which depend on $q$, from the leading terms of the $\hbar$-expansion (\ref{mmr-intro}). There are also different expansions in $\hbar$, associated to other flat connections, in particular the one that arises in the volume conjecture. All these expansions should be related by the resurgence, as discussed to some extent e.g. in \cite{Gukov:2019mnk}; it is thus important to understand how our results fit into this framework. In a similar vein, it is worth uncovering how our results are related to quantum modularity, whose role in the context of $F_K(x,q)$ invariants is discussed in \cite{Ekholm:2020lqy}. Yet another interesting question is how the expansion (\ref{mmr-intro}), and our reconstruction prescription, are constrained by the underlying integral structure of colored HOMFLY-PT polynomials (as captured by LMOV invariants), or the integral structure of corresponding quiver generating series (captured by motivic Donaldson-Thomas invariants) \cite{Kucharski:2017poe,Kucharski:2017ogk}. Furthermore, it would be rewarding to lift our reconstruction program to a homological level, and to understand what relations between knot Floer homology and colored HOMFLY-PT homology it would then reveal.

The plan of the paper is as follows. In section \ref{sec-ingredients} we concisely summarize various aspects of knot invariants that play a role in the rest of the paper. In section \ref{ssec-procedure} we present our reconstruction procedure. In section \ref{ssec-reconstruct} we illustrate how this procedure works in various examples, which involve knots up to 8 crossings and some infinite series, and in particular we derive expressions for colored superpolynomials for $7_4$ knot.


\section{An unexpected party}   \label{sec-ingredients}

In this section we briefly summarize various developments in knot theory, and related physical concepts, that play a role in what follows. At the same time, we also set up the notation that we use in the rest of the paper.


\subsection{Alexander polynomial}   \label{ssec-Alexander}

Alexander polynomial $\Delta(x)$ can be defined in various ways: by considering a knot diagram and associating weights to its crossings, in terms of a Seifert matrix, by Conway's skein relations, etc. These are standard definitions, so instead of providing details, in table \ref{tab-A} we list explicitly Alexander polynomials for various knots that we will consider in section \ref{ssec-reconstruct}. Note that $\Delta(x)=\Delta(x^{-1})$ and we use normalization $\Delta(1)=1$. Apart from the dependence on $x$, we also express $\Delta(x)$ in terms of a variable $X=\frac{(1-x)^2}{x}$, which plays a crucial role in cyclotomic expansions that we discuss in what follows.

\begin{table}
\be
\nonumber
\begin{array}{|c|c|}
\hline
\textrm{Knot} & \Delta(x) \\
\hline  
3_1 & x^{-1}-1+x = 1 + X\\ 
4_1 &-x^{-1}+ 3-x = 1 - X\\
5_1 & x^{-2}-x^{-1}+1-x+x^2 = 1 +3X+X^2\\
5_2 & 2x^{-1}-3+2x = 1 + 2X\\
6_1 & -2x^{-1}+5-2x = 1 - 2X\\
6_2 & -x^{-2}+3x^{-1}-3+3x-x^2 = 1 - X - X^2\\
6_3 & x^{-2}-3x^{-1}+5-3x+x^2 = 1 + X + X^2\\
7_1 & x^{-3}-x^{-2}+x^{-1} -1+x-x^2+x^3 = 1 + 6X+5X^2+X^3 \\
7_2 & 3x^{-1}-5+3x = 1+3X\\
7_3 & 2x^{-2}-3x^{-1}+3-3x+2x^2 = 1 +5X+2X^2 \\
7_4 & 4x^{-1}-7+4x = 1+4X\\
9_2 & 4x^{-1}-7+4x = 1+4X \\
8_{19} & x^{-3}-x^{-2}+1-x^2+x^3= 1 + 5X+5X^2+X^3 \\
\textrm{Torus knots } T^{2,2p+1}  & \sum_{i=-p}^p (-1)^{p+i}x^i = \frac{1}{x^p}\frac{1+x^{2p+1}}{1+x} =1+ \sum_{i=1}^{p}  {p+i \choose 2i}    X^i \\
\textrm{Twist knots } 4_1, 6_1, 8_1, \ldots & -px^{-1} + (1+2p)- px=1-pX \\
\textrm{Twist knots } 3_1, 5_2, 7_2, \ldots & px^{-1} + (1-2p)+ px=1+pX \\
\hline
\end{array}
\ee
\caption{Alexander polynomials $\Delta(x)$ for various knots, written as a function of $x$ and in terms of $X=\frac{(1-x)^2}{x}$.}   \label{tab-A}
\end{table} 

Furthermore, Alexander polynomial arises as an Euler characteristic of knot Floer homology $\widehat{HFK}$ \cite{manolescu2014introduction}
\be 
\Delta(q) = \sum_{d,s} (-1)^d q^s \,  {\mathrm {dim}}\, \widehat{HFK}_d (s).
\ee 
Knot Floer homology is expected to arise through the action of a certain differential from HOMFLY-PT homology that we briefly discuss in the next section. Also the Melvin-Morton-Rozansky expansion (\ref{mmr-intro}), which relates Alexander polynomial to colored HOMLFY-PT polynomials, indicates on decategorified level that a relation between corresponding homology theories should exist. From the homological perspective, it is also natural to consider the Poincar\'{e} characteristic of $\widehat{HFK}$
\be 
{\mathrm{HFK}} (q,t) = \sum_{d,s} t^d q^s \,  {\mathrm {dim}}\, \HFK_d (s).    \label{HFK}
\ee 
This provides a $t$-deformation of Alexander polynomial, and can be also obtained as a specialization of a superpolynomial, i.e. Poincar\'{e} characteristic of HOMFLY-PT homology, briefly discussed in the next section.


\subsection{HOMFLY-PT homology}    \label{ssec-homology}

While knot Floer homology categorifies Alexander polynomial, it is expected that there exists a triply-graded HOMFLY-PT homology $\mathcal{H}_{i,j,k}$ that categorifies HOMFLY-PT polynomial $P(a,q)$, so that 
\be
P(a,q) = \sum_{i,j,k} a^i q^j (-1)^k \dim \mathcal{H}_{i,j,k}.   \label{homfly}
\ee
An explicit construction of the homology theory $\mathcal{H}_{i,j,k}$ is not known, however it is expected to satisfy a number of properties, which include the existence of certain differentials $d_N$ for $N\in\mathbb{Z}$  \cite{Dunfield:2005si}. These properties impose constraints that enable to determine e.g. a superpolynomial, i.e. Poincar\'{e} characteristic of $\mathcal{H}_{i,j,k}$
\be 
P(a,q,t) = \sum_{i,j,k} a^i q^j t^k \dim \mathcal{H}_{i,j,k} = \sum_i a^{a_i}q^{q_i} t^{t_i},   \label{superpolynomial}
\ee 
where $i$ in the summation in the expression on the right labels homology generators, so that each monomial in the superpolynomial represents one generator, and $(a_i,q_i,t_i)$ are referred to as degrees of the $i$'th generator. Conjecturally, the action of $d_N$ differentials with $N>0$ reduces the HOMFLY-PT homology to the $sl_N$ Khovanov-Rozansky homology, while the action of $d_0$ reduces the HOMFLY-PT homology to the knot Floer homology. In consequence, HOMFLY-PT and Alexander polynomials arise as specializations of the superpolynomial,
$P(a,q) = P(a,q,-1)$ and $\Delta(q) = P(1,q,-1)$. Superpolynomials for various knots are listed in table \ref{tab-P}.

An important quantity that characterizes generators of HOMFLY-PT homology is the so-called $\delta$-grading \cite{Dunfield:2005si}, defined for the $i$'th generator as
\be
\delta_i = 2a_i+q_i-t_i.  \label{delta-grading}
\ee
A knot is called thin if all its generators have the same $\delta$-grading, $\delta=\delta_i=const$. Knots that do not satisfy this condition are called thick. 

\begin{table}
\be
\nonumber
\begin{array}{|c|c|}
\hline
\textrm{Knot} &  P_{\square}(a,q,t)\\
\hline  
3_1 & \frac{a}{q} +a^2t^3+ aqt^2 \\  
4_1 & \frac{1}{qt}+1+\frac{1}{at^2}+at^2+q t\\  
5_1 & \frac{a^2}{q^2}+\frac{a^3t^3}{q} +a^2t^2+a^3qt^5+ a^2q^2t^4\\  
5_2 & \frac{a}{q}+\frac{a^2t^2}{q}+at+a^2t^3+a^3t^5+a^2qt^4+aqt^2\\  
6_1 & \frac{1}{qt}+\frac{at}{q} +\frac{1}{at^2}+2   +  at^2+a^2t^4 + qt+aqt^3\\   
6_2 & \frac{a}{q^2t}+ \frac{1}{qt^2}+ \frac{a}{q}+ \frac{a^2t^2}{q} +2at+a^2t^3+ q+aqt^2+a^2qt^4+aq^2t^3 \\   
6_3 & \frac{1}{q^2t^2}+\frac{1}{aqt^3}+\frac{1}{qt} +\frac{at}{q}+\frac{1}{at^2}+3+at^2+\frac{q}{at}+qt+aqt^3 +q^2t^2  \\    
7_1 & \frac{a^3}{q^3} + \frac{a^4t^3}{q^2}+\frac{a^3t^2}{q}+ a^4t^5  +a^3qt^4+a^4q^2t^7 +a^3q^3t^6 \\   
7_2 & \frac{a}{q}+\frac{a^2t^2}{q}+\frac{a^3t^4}{q}+at+2a^2t^3 + a^3t^5+a^4t^7+ aqt^2 +a^2qt^4+a^3qt^6  \\   
7_3 & \frac{1}{a^{3}q^{2}t^{6}}+\frac{1}{a^{2}q^{2}t^{4}}  + \frac{1}{a^{4}qt^{7}} + \frac{1}{a^{3}qt^{5}} +\frac{1}{a^{3}qt^{3}}+\frac{2}{a^{3}t^{4}}+\frac{1}{a^{2}t^{2}} + \frac{q}{a^{4}t^{5}} + \frac{q}{a^{3}t^{3}}+\frac{q}{a^{2}t} +\frac{q^2}{a^{3}t^{2}}+\frac{q^2}{a^2}\\   
7_4 &\frac{a}{q} +\frac{2a^2t^2}{q}+ \frac{a^3t^4}{q}+ 2at +2a^2t^3 +2a^3t^5+ a^4t^7+aqt^2+2a^2qt^4 +a^3qt^6\\   
9_2 &  \frac{a}{q} + \frac{a^2t^2}{q}  +\frac{a^3t^4}{q} +\frac{a^4t^6}{q}+at+2a^2t^3 +2a^3t^5+a^4t^7 +a^5t^9+ \\
 &  + aqt^2+   a^2qt^4+ a^3qt^6+a^4qt^8 \\
8_{19} & \frac{a^3}{q^{3}}+\frac{a^4 t^3}{q^2}+\frac{a^3 t^2}{q}+\frac{a^4 t^5}{q}+a^3t^4+a^4t^5+a^5t^8+a^3qt^4+a^4qt^7 + a^4q^2t^7+a^3q^3t^6 \\   
T^{2,2p+1}  & \frac{(a/q)^p}{1-q^2t^2}\big(1-(qt)^{2p+2}+aqt^3(1-(qt)^{2p})\big) \\   
4_1,6_1,8_1,\ldots & 1+(1+atq^{-1})(1+aqt^3)   (a^{-1}t^{-2}+1+\ldots+a^{p-2}t^{2p-4}) \\   
3_1,5_2,7_2,\ldots & -t^{-1} + (1+atq^{-1})(1+aqt^3)  (t^{-1}+at +\ldots+a^{p-1}t^{2p-3})  \\
\hline
\end{array}
\ee
\caption{(Uncolored) superpolynomials $P(a,q,t)\equiv P_\square(a,q,t)$ for various knots.}  \label{tab-P}
\end{table} 

\begin{figure}[htp]
    \centering
    \includegraphics[width=0.4\textwidth]{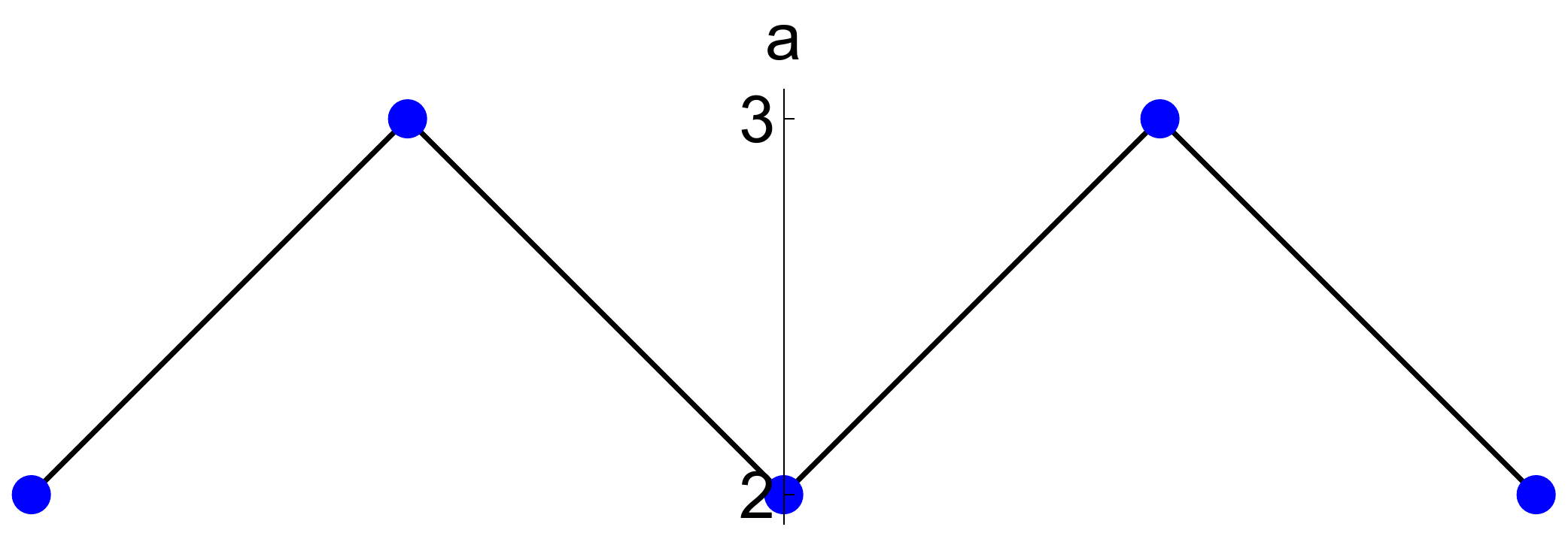}
    \caption{Homological diagram for $5_1$ knot.}\label{fig-51}
\end{figure}

It is useful to present the structure of a superpolynomial in a so-called homological diagram, whose horizontal and vertical axes encode $q$-degree and $a$-degree of generators respectively. Each dot in a diagram represents one generator or the corresponding monomial in the superpolynomial. Examples of homological diagrams are shown in figures \ref{fig-31}, \ref{fig-41}, \ref{fig-51}, etc. 

One aspect of the HOMFLY-PT homology that we take advantage of, are relations between its generators imposed by the canceling differentials \cite{Dunfield:2005si}. In particular it follows that generators of the HOMFLY-PT homology can be assembled into two types of structures, which we refer to as a zig-zag and a diamond. We denote zig-zags and diamonds in all homological diagrams presented in what follows. A given diagram necessarily contains one zig-zag, which consists of an odd number of dots, and a number of diamonds. 

Conjecturally, there exists also colored HOMFLY-PT homology $\mathcal{H}^R_{i,j,k}$, which categorifies HOMFLY-PT polynomials colored by a representation $R$
\be
P_R(a,q) = \sum_{i,j,k} a^i q^j (-1)^k \dim \mathcal{H}^{R}_{i,j,k}.   \label{colored-homfly}
\ee
The structure and properties of various differentials in such homology theories, for symmetric representations $R=S^r$, have been formulated in \cite{Gukov:2011ry}. For a large class of knots, these structural properties enable to determine corresponding colored superpolynomials, defined as
\be 
P_r(a,q,t) = \sum_{i,j,k} a^i q^j t^k \dim \mathcal{H}^{S^r}_{i,j,k}.   \label{Pr-aqt}
\ee 
For example, such colored homology theories are expected to possess canceling differentials, which impose the following conditions on colored superpolynomials
\be
\begin{split}
P_r(a,q,t)&=a^{-rs}q^{r s}t^0+(1+aq^{-1}t)Q_1(a,q,t)   = \\
&= a^{-rs}q^{-r^2s}t^{-2rs}+(1+aq^{r}t^3)Q_r(a,q,t),
\label{eq:Cancelling}
\end{split}
\ee
where $Q_1$ and $Q_r$ are polynomials in $a,q$ and $t$ with positive coefficients, and $s$ is the Rasmussen invariant. These conditions, and other analogous conditions arising from the existence of other differentials, strongly constrain the form of superpolynomials. In what follows we  verify whether colored HOMFY-PT polynomials or superpolynomials that we determine for various knots indeed satisfy these conditions. For brevity, and analogously as in (\ref{Pr-aqt}), we denote HOMFLY-PT polynomials colored by symmetric representations $R=S^r$ simply by
\be 
P_r(a,q) = P_r(a,q,-1) = \sum_{i,j,k} a^i q^j (-1)^k \dim \mathcal{H}^{S^r}_{i,j,k}.   \label{colored-homfly-r}
\ee 

For $r=1$, corresponding knot polynomials (Jones, HOMFLY-PT (\ref{homfly}), superpolynomial (\ref{superpolynomial})) are referred to as uncolored, and denoted $P_1(q)\equiv  P(q)\equiv P_\square(q)$ (where we skip a dependence on other variables, such as $a$ or $t$). Recall that some knots satisfy the exponential growth condition, which is the statement that their colored superpolynomials (or various specializations thereof) satisfy the relation
\be
P_r(a,1,t)=P_{\square}(a,1,t)^r.   \label{exp-growth}
\ee
We also verify whether the expressions that we find satisfy this condition, if only it is expected to hold for the knot under consideration.

In addition to the above 3 gradings, it is conjectured that also a closely related quadruply-graded homology theory and corresponding superpolynomials exist \cite{Gorsky:2013jxa}. It is not hard to generalize all our results to such a quadruply-graded theory, however we skip this task for brevity.


\subsection{Cyclotomic expansions}    \label{ssec-cyclo}

Colored knot polynomials $P_r(q)\equiv P_r^K(q)$ for a knot $K$ (here we skip dependence on other variables, such as $a$ or $t$), colored by symmetric representations $S^r$, are expected to have cyclotomic expansions, which is the statement that they can be written in the form
\be
P_r^K(q) = \sum_{m=0}^r c_m(r,q)  d^K_m(q),
\ee
where $c_m(r,q)$ is a universal factor that captures the whole dependence on color $r$ and does not depend on a knot $K$, while color-independent factor $d^K_m(q)$ depends on a choice of a knot $K$. In case of Jones polynomial the existence of such an expansion is proven in \cite{Habiro_2007}; in this case $c_m(r,q)=(q^{r+2};q)_m (q^{-r};q)_m$, for $(x;q)_m=\prod_{i=0}^{m-1}(1-xq^i)$, so that 
\be
J^K_r(q) = \sum_{m=0}^r   (q^{r+2};q)_m (q^{-r};q)_m  d^K_m(q).
\ee
Subsequently, analogous conjectures were formulated for other colored knot polynomials: for refined (or generalized) Jones polynomial in \cite{berest2019cyclotomic}, and for HOMFLY-PT polynomial and superpolynomial in \cite{Nawata:2015wya}. For colored HOMFLY-PT polynomial, the cyclotomic expansion is characterized by $c_m(r,q) =a^{\alpha r}q^{\beta r} {r \brack m} q^{-rm} (aq^r;q)_m$, so that
\be
P^K_r(a,q) = a^{\alpha r}q^{\beta r} \sum_{m=0}^r   {r \brack m} q^{-rm} (aq^r;q)_m  d^K_m(a,q),  \label{cyclo}
\ee
for some $\alpha$ and $\beta$. A generalization of this statement to superpolynomials takes form \cite{Nawata:2015wya}
\be
P^K_r(a,q,t) = \pm a^{\alpha r}q^{\beta r}t^{\gamma r} \sum_{m=0}^r   {r \brack m} q^{-rm} (-aq^rt^3;q)_m  e^K_m(a,q,t),   \label{cyclo-t}
\ee
for some $\alpha,\beta$ and $\gamma$.


\subsection{Knots-quivers correspondence}    \label{ssec-quivers}

The last ingredient playing an important role in our analysis is the knots-quivers correspondence, found in \cite{Kucharski:2017poe,Kucharski:2017ogk}. According to this correspondence, colored HOMFLY-PT homology, and thus all knot polynomials mentioned above too, are related to quiver representation theory. Namely, it turns out that generators of HOMFLY-PT homology for a given knot, which in string theory realization are identified with certain fundamental BPS states, can be also associated to nodes of some particular quiver. There are $C_{i,j}$ arrows between the $i$'th and the $j$'th node in this quiver, which encode interactions between these BPS states. To such a quiver one can assign the so-called quiver generating series \cite{Kontsevich:2010px,reineke2011degenerate}, defined as $P(x_1,\ldots,x_m) = \sum_{r=0}^{\infty} P_r(x_1,\ldots,x_m)$, where
\be
P_r(x_1,\ldots,x_m) = \sum_{d_1+\ldots +d_m=r}  \frac{q^{\sum_{i,j} C_{i,j} d_i d_j} (q;q)_r}{\prod_{i=1}^m (q;q)_{d_i}}  x_1^{d_1}\cdots x_m^{d_m}.   \label{Pr-quiver}
\ee
As found in \cite{Kucharski:2017poe,Kucharski:2017ogk}, this generating series is equal to the generating function of colored HOMFLY-PT polynomials, with (\ref{Pr-quiver}) specializing to (\ref{colored-homfly-r}), upon the identification $x_i = q^{q_i-t_i} a^{a_i} (-1)^{t_i}$, where $a_i, q_i$ and $t_i$ are powers of $a,q$ and $t$ in various monomials in the uncolored superpolynomial (\ref{superpolynomial}). This explains the relation of quivers to HOMFLY-PT homology. Furthermore, factorization of the above series determines LMOV invariants, i.e. multiplicities of composite BPS states, which are made of the above mentioned fundamental BPS states. A simple generalization of the above expression enables also to present the generating function of colored superpolynomials (\ref{Pr-aqt}) in the form of an analogous quiver generating series, determined by the same quiver, whose structure is captured by the numbers of arrows $C_{i,j}$ \cite{Kucharski:2017ogk}. 

Quivers corresponding to a large class of knots were found in \cite{Kucharski:2017ogk,Panfil:2018sis}; more recently they were also identified for all rational knots \cite{Stosic:2017wno} and arborescent knots \cite{Stosic:2020xwn}. Geometric interpretation of knots-quivers is discussed in \cite{Ekholm:2018eee,Ekholm:2019lmb}, and its generalization in the context of topological string theory in \cite{Panfil:2018faz}.

In what follows we also identify quivers corresponding to various knots, by rewriting colored polynomials in the form (\ref{Pr-quiver}) with the above mentioned identification of $x_i$. To this end we take advantage, among the others, of the following relation
\begin{equation}
\label{identity-qpoch}
\frac{(a;q)_{d_1+d_2+\ldots+d_k}}{\prod_{i=1}^k (q;q)_{d_i}}    = \, \sum_{\substack{\alpha_i+\beta_i=d_i\\ i=1,\ldots,k}}
\frac{(-a)^{\sum_{i=1}^k \alpha_i} q^{\frac12\sum_{i=1}^k (\alpha_i^2 -\alpha_i)}
q^{\sum_{i=1}^{k-1}(\alpha_{i+1}\sum_{j=1}^i d_j)}}{\prod_{i=1}^k (q;q)_{\alpha_i} (q;q)_{\beta_i}}.
\end{equation}


\section{Back again from Alexander polynomial}    \label{ssec-procedure}

In this section we present the main idea of this work, i.e. we explain how to reconstruct colored HOMFLY-PT polynomials and superpolynomials, as well as $F_K(x,a,q)$ invariants for some knot complements, from Alexander polynomial, by means of the Melvin-Morton-Rozansky expansion (\ref{mmr-intro}), and making contact with various concepts presented in section \ref{sec-ingredients}. The first step in this process is appropriate rewriting of Alexander polynomial, based on the inverse binomial theorem. Depending on a choice of such rewriting, we obtain various forms of final expressions. The first rewriting, that we refer to as ``homological'', is discussed in section \ref{ssec-1st}; it relies on the structure of superpolynomials and HOMFLY-PT homology and leads to expressions for colored polynomials which are most appropriate from the perspective of knots-quivers correspondence. Another rewriting, which we call ``cyclotomic'', is presented in section \ref{ssec-2nd}, and its advantage is an immediate connection to cyclotomic expansions. In section \ref{ssec-FK} we discuss the relation of these results to $F_K(x,a,q)$ invariants.


\subsection{The ``homological'' approach}   \label{ssec-1st}

Our starting point is Alexander polynomial $\Delta(x)$. In the first, ``homological'' approach, we write it in a way that makes manifest the structure of generators of the uncolored HOMFLY-PT homology. As we reviewed in section \ref{ssec-homology}, for each knot these generators are assembled into one zig-zag and several diamonds. All these generators correspond to monomials in the superpolynomial, so the number of such monomials is equal to the number of generators. Moreover, Alexander polynomial arises as a specialization $a=1=-t$ of the superpolynomial, i.e. $\Delta(q) = P_{\square}(a=1,q,t=-1)$, so clearly monomials in Alexander polynomial are related to those in superpolynomial, possibly up to some cancellations that might arise upon the specialization $t=-1$. Let us show first, that such cancellations in fact do not arise for thin knots.

Thus, consider the superpolynomial $P_{\square}(a,q,t)=\sum_{i} a^{a_i} q^{q_i} t^{t_i}$ for a knot that is thin, so that $\delta = 2a_i+q_i-t_i$ takes the same value for each generator $i$ (i.e. for each term in the above summation). It follows that the combination $q_i-t_i$ has the same parity for all generators (since $2a_i$ is even). Furthermore, if a pair of generators would cancel due to a minus sign that would arise upon $t=-1$ specialization to Alexander polynomial, their $t_i$'s would have different parity. It follows that their $q_i$'s would also have different parity -- so powers of $q$ in those terms would be different, and thus they could not cancel. It follows that for thin knots the number of monomials does not change upon the specialization $a=1=-t$ from superpolynomial to Alexander polynomial, and therefore each monomial in Alexander polynomial corresponds to a particular homology generator. This also implies that the sum of absolute values of coefficients in Alexander polynomial is equal to the number of monomials in the superpolynomial, which is equal to $P_{\square}(1,1,1)$.

To sum up, for thin knots, monomials in Alexander polynomial can be immediately grouped into a zig-zag and diamonds, analogously as in a superpolynomial. For thick knots one can also group the terms in Alexander polynomial into such patterns, however to this end some additional terms must be first added and subtracted (those which cancel upon reduction of a superpolynomial to Alexander polynomial). Nonetheless, at least for relatively simple thick knots, it is also not hard to determine how to split Alexander polynomial into terms that form a zig-zag and diamonds. In section \ref{ssec-819} we demonstrate that our reconstruction procedure works for $8_{19}$ knot, which is thick.

Therefore, suppose that in Alexander polynomial $\Delta(x)$ we identified terms associated to a zig-zag, which we denote $\Delta_z(x)$, while the terms that correspond to diamonds we denote $\Delta_i(x)$ (for various diamonds labeled by $i$). Suppose that the zig-zag has length $2p+1$. It follows from specialization of a superpolynomial that it must have form
\be
\Delta_z(x) = x^{-p} - x^{-p+1} +\ldots -x^{p-1} + x^p = x^{-p}\big(1 - x(1-x) \sum_{j=0}^{p-1} x^{2j}\big),  \label{zigzag-p}
\ee
where interchanging signs arise from properties of canceling differentials, while the overall sign follows from the normalization $\Delta(1)=1$ (because contributions from diamonds cancel for $x=1$ and do not affect this overall sign). Furthermore, from specialization of a superpolynomial and properties of canceling differentials it follows that terms assembled into diamonds have form
\be
\Delta_i(x) = (-1)^{k_i} x^{s_i} (1-x)^2,   \label{diamond-i}
\ee
for some particular $k_i$ and $s_i$ for the $i$'th diamond. Altogether, it follows that
\be
\Delta(x) = \Delta_z(x) + \sum_i \Delta_i(x) = x^{-p}\big(1 - (1-x)f(x)\big),   \label{Delta-f}
\ee 
where
\be
f(x) = x\sum_{j=0}^{p-1} x^{2j} - (1-x) \sum_i (-1)^{k_i} x^{p+s_i} .   \label{fx}
\ee

Note that the first term $1$ in the large bracket in (\ref{Delta-f}) represents the left-most end of a zig-zag. 
We can now use the inverse binomial theorem 
\be
\frac{1}{(1- u)^n} = \sum_{m=0}^{\infty} { n+m-1 \choose m} u^m.   \label{inverse-binomial}
\ee
Identifying $u=(1-x)f(x)$, we write the leading term in the Melvin-Morton-Rozansky expansion (\ref{mmr-intro}) as a series
\be
\frac{1}{\Delta(x)^{N-1}} = x^{p(N-1)} \sum_{m=0}^{\infty} {N+m-2 \choose m} (1-x)^m f(x)^m. \label{inverse-Delta}
\ee
Moreover, from the subsequent multinomial expansion of $f(x)^m$, we get additional binomial coefficients. For example, if $f(x) = x\sum_{j=0}^{p-1} x^{2j}$, i.e. we have only one zig-zag and no diamonds, we get
\be
\Big(x\sum_{j=0}^{p-1} x^{2j}\Big)^m = \sum_{0\leq k_{p-1} \leq k_{p-2} \leq \ldots \leq k_{1} \leq m} {m \choose k_1}{k_1 \choose k_2}\cdots{k_{p-2} \choose k_{p-1}} x^{m+2(k_1 + \ldots + k_{p-1})}.  \label{zigzag-binomials}
\ee
If in addition to a zig-zag there are some diamonds, we still get similar expressions, which involve a number of binomial coefficients.

Therefore the series (\ref{inverse-Delta}), with additional expansion of the term $f(x)^m$ as in (or generalizing) (\ref{zigzag-binomials}), is the expression that we wish to promote to colored HOMFLY-PT polynomials, superpolynomials, or hopfully to $F_K(x,a,q)$ invariants, by appropriate $q$-, $a$-, and $t$-deformation. Such deformations can be implemented by invoking various features presented in section \ref{sec-ingredients}, and comparing either with several first colored polynomials, or several first coefficients $R_k(x,N)$ in (\ref{mmr-intro}), in case they are known independently. As a general strategy, we replace binomials by $q$-binomials, replace Pochhammer symbols by $q$-Pochahmmer symbols, replace powers of certain expressions by $q$-Pochhammers, and allow introducing in the summand extra overall powers of $a$ and $t$ that are linear in summation variables, and extra overall powers of $q$ that are quadratic in summation variables. In what follows we use the following notations for the Pochhammer and $q$-Pochhammers symbols
\be
(k)_m=\prod_{i=0}^{m-1} (k-i), \qquad\quad (y;q)_m = \prod_{i=0}^{m-1} (1 - yq^i),
\ee
and the $q$-binomials are defined as
\be
\qquad {r \brack m} = \frac{(q;q)_r}{(q,q)_m (q,q)_{r-m}} = \frac{(q^r;q^{-1})_m}{(q;q)_m}.
\ee

The above general strategy concerns in particular the terms ${N+m-2 \choose m} (1-x)^m =  \frac{(N-1)_m}{m!} (1-x)^m$ that arise universally in (\ref{inverse-Delta}). To get HOMFLY-PT polynomials, it is natural to deform them as follows
\be
\begin{split}
{N+m-2 \choose m} (1-x)^m   \ \rightsquigarrow\   \frac{(q^{N-1};q)_m }{(q;q)_m} (x;q^{-1})_m
& =\frac{(aq^{-1};q)_m (x;q^{-1})_m}{(q;q)_m}  =  \\
& =  {r \brack m} (aq^{-1};q)_m
    \label{deform-binom}
\end{split}
\ee
where we identified $x=q^r$ and $a=q^N$. To get superpolynomials, the above deformation is further modified by a single factor of $t$
\be
{N+m-2 \choose m} (1-x)^m   \ \rightsquigarrow\    {r \brack m} (-aq^{-1}t;q)_m.
    \label{deform-binom-t}
\ee
The combination ${r \brack m} (-aq^{-1}t;q)_m$ indeed arises in various expressions for colored superpolynomials identified before \cite{Fuji:2012pi,Nawata:2012pg,Nawata:2015wya}. Similarly, we replace all binomials in (\ref{zigzag-binomials}) or analogous expressions by $q$-binomials. Furthermore, another factor of $(1-x)^m$ gets deformed into $(-aq^rt^3;q)_m$ that often accompanies (\ref{deform-binom-t}).

To sum up, in view of (\ref{inverse-Delta}) and (\ref{deform-binom-t}), and subsequent remarks, we predict that colored superpolynomials have the structure 
\be
P_r(a,q,t) =  \frac{a^{rp}}{q^{rp}}  \sum_{m=0}^r {r \brack m} (-aq^{-1}t;q)_m  f_q(q^r,a,q,t)^m,   \label{Pr-q-deform}
\ee
where $f_q(q^r,a,q,t)^m$ is a deformation of $f(x)^m$, such that binomials are replaced by $q$-binomials, and the summand in addition involves $q$ raised to a power that is at most quadratic in summation variables, and $a$ and $t$ are raised to a linear power is summation variables. Colored HOMFLY-PT polynomials arise as $t=-1$ specialization of (\ref{Pr-q-deform}). Quadratic and linear powers mentioned above can be fixed by comparing with the first few colored superpolynomials, or HOMFLY-PT polynomials, or (in principle) with the first few coefficients $R(x,N)$ in (\ref{mmr-intro}). We also verify that resulting expressions are consistent with differentials in HOMFLY-PT homologies and satisfy conditions such as (\ref{eq:Cancelling}), and (when relevant) with the exponential growth (\ref{exp-growth}).  


\subsection{The ``cyclotomic'' approach}   \label{ssec-2nd}

Let us present now another expansion that we consider. It leads to expressions that are equivalent to those found as in the previous section, however now they are written in the cyclotomic form. To this aim, the main observation is that Alexander polynomial can be written as a polynomial in $X=\frac{(1-x)^2}{x}$, in the form
\be
\Delta(x) = 1 - \sum_{i=1}^s a_i \frac{(1-x)^{2i}}{x^i} \equiv 1 - g(X).   \label{Delta-d}
\ee
Indeed, we know that $\Delta(x)=\Delta(-x)$, so $\Delta(x)$ clearly can be written as a combination of powers of $\frac{1}{x}-2+x=\frac{(1-x)^2}{x}$. One can adjust first the coefficient $a_s$ by comparing with the term at the highest power of $x$ in $\Delta(x)$, then adjust $a_{s-1}$ by comparing with the next-to-highest power $x^{s-1}$, etc. At the end a constant term needs to be fixed. However, we also know that $\Delta(1)=1$, and for $x=1$ the sum over $i$ in (\ref{Delta-d}) vanishes, so the remaining constant term is $1$, and altogether we obtain (\ref{Delta-d}). Examples of Alexander polynomials written in terms of $X$ are provided in table \ref{tab-A}.

The above statement can be also related to the structure of a zig-zag and diamonds in (\ref{Delta-f}). Namely, one can complete a zig-zag to a combination of diamonds, with an extra operation that removes one corner of one diamond. The simplest example of such a process arises for a zig-zag of length 3, for which we can write
\be
x^{-1}-1+x=x^{-1}\big(1-x(1-x)\big) = 1+\frac{(1-x)^2}{x}.    \label{zigzag-3}
\ee
The middle expression above is written in the form (\ref{zigzag-p}) for $p=1$, while the form on the right can be interpreted as a diamond with one corner (represented by the first ``1'') removed. For longer zig-zags such an interpretation works analogously. It then follows that each zig-zag can be written in the form $\Delta_z(x) =1 + h(X)$ for some polynomial $h(X)$, and also each diamond has the structure as in (\ref{diamond-i}). Therefore the whole $\Delta(x)$ can be written in the form (\ref{Delta-d}) (which is still not entirely obvious due to overall powers of $x$ at each diamond (\ref{diamond-i})). Nonetheless, in this ``cyclotomic'' approach, these structural features are not that essential, and it is simply crucial that we can write Alexander polynomial in the form (\ref{Delta-d}).

Thus, starting now with (\ref{Delta-d}), we make analogous inverse binomial expansion as in the previous section. Note that the interpretation of the term $1$ that enables the expansion is different -- as mentioned above, it represents a corner of a certain virtual diamond, obtained from filling in a zig-zag. The inverse binomial theorem (\ref{inverse-binomial}) now yields
\be
\frac{1}{\Delta(x)^{N-1}} = \sum_{m=0}^{\infty} {N+m-2 \choose m}  g(X)^m =  \sum_{k=0}^{\infty} c_k X^k. \label{inverse-Delta2}
\ee
Note that in comparison to (\ref{inverse-Delta}) there is no overall term $x^{p(N-1)}$, and most importantly the whole dependence on $x$ arises only through $X$. This enables us, after appropriate expansion of $g(X)^m$ and rearranging summations, to write the above expression as a series in $X$ with certain coefficients $c_k$. 

Having found the expansion (\ref{inverse-Delta2}), in the second step we can deform it. Regarding $X^k=\frac{(1-x)^{2k}}{x^k}$, we find that in general one factor of $(1-x)^k$ is coupled to some other $k!$ in denominator and gives rise to ${r \brack k}$, analogously as in (\ref{deform-binom}). Moreover, we still have the second factor of $(1-x)^k$, which universally gets deformed to
\be
(1-x)^k \ \rightsquigarrow\  (-aq^r t^3;q)_k,   \label{aqrk}
\ee
while $x^k$ in the denominator is simply $q^{-kr}$. Altogether these terms combine to 
\be
\frac{(1-x)^{2k}}{k!x^k} \ \rightsquigarrow\   {r \brack k}  q^{-rk} (-aq^r t^3;q)_k ,   \label{deform-cyclo}
\ee
which captures the whole dependence on $x=q^r$, and these are precisely the factors that appear in the cyclotomic expansion (\ref{cyclo-t}) of superpolynomials, or (\ref{cyclo}) for HOMFLY-PT polynomials. Furthermore, in coefficients $k! c_k$ we replace Pochhammers and binomials by $q$-Pochhammers and $q$-binomials, and allow an additional deformation that may involve at most powers of $a$ and $t$ linear in summation variables and powers of $q$ quadratic in summation variables. Ultimately we get
\be
P_r(a,q,t) =   \sum_{k=0}^r {r \brack k} q^{-rk} (-aq^rt^3;q)_k   \widetilde{c}_k,   \label{Pr-q-deform2}
\ee
where $\widetilde{c}_k=\widetilde{c}_k(a,q,t)$ is a deformation $k! c_k$. This is how a cyclotomic expansion of colored polynomials arises from this second form of expansion of Alexander polynomial.


\subsection{Relation to $F_K(x,a,q)$ invariants}   \label{ssec-FK}

Let us also discuss a possible relation to $F_K(x,q)$ or $F_K(x,a,q)$ invariants, introduced recently in \cite{Gukov:2019mnk,Park:2019xey,Ekholm:2020lqy}. These invariants could be derived analogously as above, if only it would be possible to write the resulting series as a well-defined expansion in $x$. Considering the first, ``homological'' approach, and taking advantage of the first line of (\ref{deform-binom}), and adjusting normalization by removing the overall prefactor $a^{rp}q^{-rp}$ (as in \cite{Ekholm:2020lqy}), one might hope to get
\be
F_K(x,a,q) = \sum_{m=0}^{\infty} \frac{(aq^{-1};q)_m (x;q^{-1})_m}{(q;q)_m}  f_q(x,a,q)^m.  \label{FK-xaq}
\ee
In particular, note that the whole dependence on $x$ follows from (\ref{inverse-Delta}), while subsequent deformations are $x$-independent. Note however, that this expression is a well-defined series in $x$, if $x$ appears only in positive powers in (\ref{fx}). This is so for $(2,2p+1)$ torus knots, whose homological diagram consists of a zig-zag only, which is represented by the first summation in (\ref{fx}). For other knots, whose diagrams involve at least one diamond, the second summation in (\ref{fx}) may introduce negative powers of $x$ due to negative values of $s_i$. This is indeed the case for other examples that we discuss in this paper, and thus for them we cannot identify $F_K(x,a,q)$ in this way. It is an interesting problem whether there exist knots for which all powers of $x$ in (\ref{fx}) are positive -- for such knots, $F_K(x,a,q)$ invariants should be given by the above formula. On the other hand, one might hope to identify some analytic continuation of the above expression to positive powers of $x$, that would lead to proper $F_K(x,a,q)$ invariants for other knots too.

Also, note that the second, ``cyclotomic'' approach, does not seem to be relevant for identifying $F_K(x,a,q)$ invariants. In this case the formula (\ref{Pr-q-deform2}) contains explicitly both positive and negative powers of $x=q^r$ under the summation, so (after extending the range of summation to infinity) we do not obtain a well defined series in $x$.


\section{Reconstructing colored (super)polynomials}    \label{ssec-reconstruct}

In this section we illustrate our reconstruction procedure for various knots up to 8 crossings, as well as for infinite series of torus knots and twist knots. We obtain expressions for colored HOMFLY-PT polynomials and superpolynomials, as well as $F_K$ invariants in the case of torus knots. Our expressions are consistent with earlier results obtained in literature, whenever they are known. Note that this asserts that these results are also consistent with the Melvin-Morton-Rozansky conjecture, which has not been verified before, and which thus provides a new independent check of their validity. 

We discuss different forms of expansion, following either the approach presented in section \ref{ssec-1st} or in section \ref{ssec-2nd}. In particular, we start the presentation by discussing $3_1$ and $4_1$ knots, not just because these are the simplest examples, but because they illustrate how to deal with basic pieces of Alexander polynomial corresponding to a zig-zag or a diamond. Indeed, a homological diagram for $3_1$ knot consists only of the shortest non-trivial zig-zag (of length 3) and no diamond. On the other hand, a diagram for $4_1$ knot consists of one diamond and a minimal zig-zag of length 1, which is represented by ``1'' in (\ref{zigzag-p}) or (\ref{Delta-d}) that is involved in the inverse binomial expansion; in consequence the function $f(x)$ in (\ref{fx}) -- that is subsequently deformed in (\ref{Pr-q-deform}) -- or $g(x)$ in (\ref{Delta-d}), involve only a contribution from one diamond. Analysis for other knots, whose diagrams involve longer zig-zags or more diamonds, essentially generalizes these two prototype cases.

Note that among other examples, we also analyze $7_4$ knot. This example is particularly interesting, because $7_4$ knot has the same Alexander polynomials as $9_2$, so that one can clearly see differences that lead to their colored superpolynomials. Moreover, HOMFLY-PT polynomials and superpolynomials for $7_4$ knot, colored by arbitrary symmetric representations, have not been determined before, so this analysis also illustrates the power of our formalism and contributes new explicit results of general interest.

Finally, to conclude, we analyze $8_{19}$ knot, i.e. $(3,4)$ torus knot, which is thick. It is reassuring to confirm explicitly that our reconstruction formalism works for thick knots too.


\subsection{Trefoil knot, $3_1$}    \label{ssec-31}

\begin{figure}[htp]
    \centering
    \includegraphics[width=0.35\textwidth]{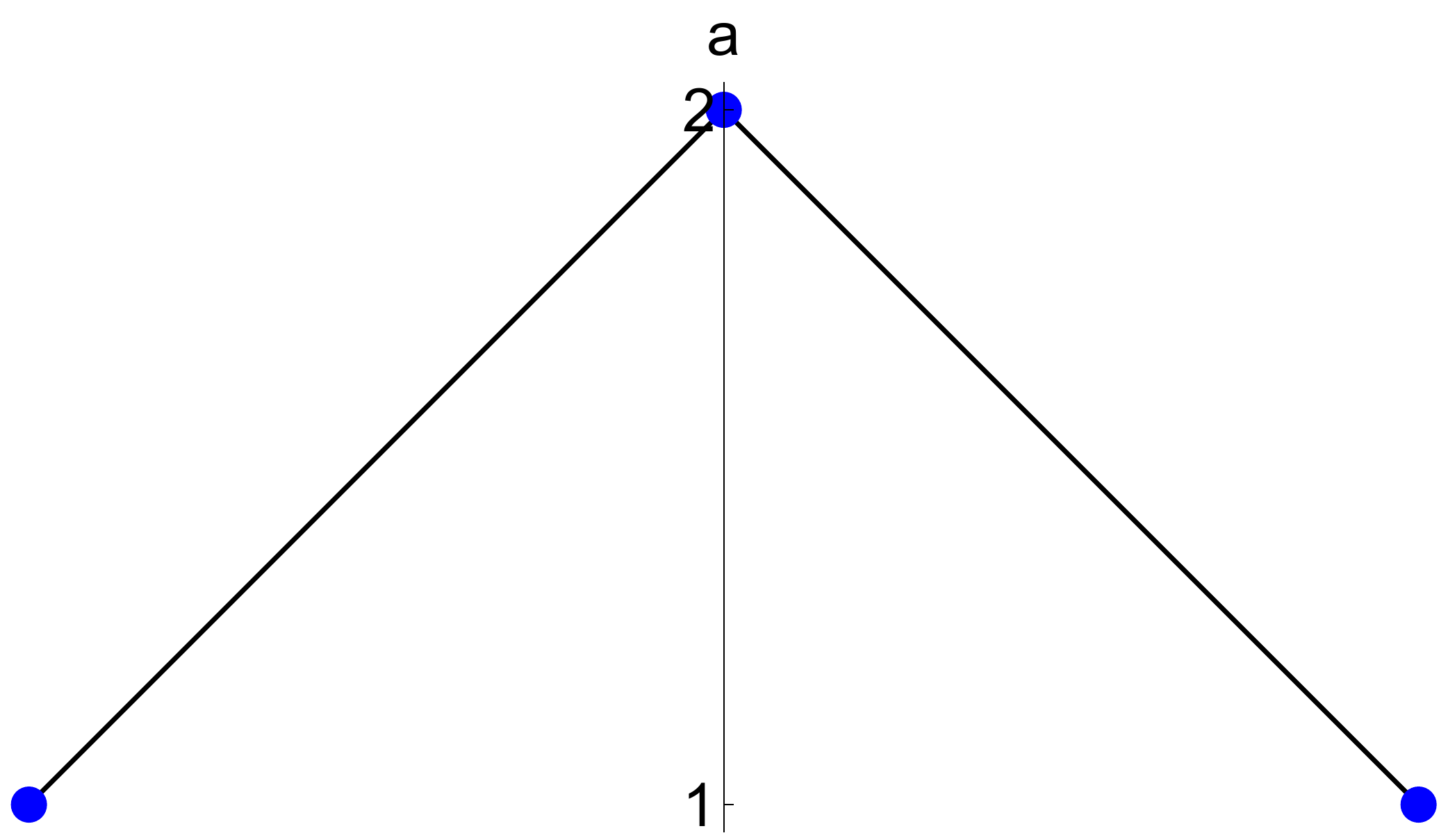}
    \caption{Homological diagram for $3_1$ knot. Each dot represents one homology generator or a monomial in the superpolynomial. Horizontal and vertical axes encode respectively $q$-degrees and $a$-degrees of generators.}\label{fig-31}
\end{figure}

In this simplest example we discuss two types of expansions, ``homological'' and ``cyclotomic'', which lead to expressions for colored polynomials that are of course equal, but have different form. The homological diagram for $3_1$ consists of a single zig-zag of length 3, see fig. \ref{fig-31}, so as we already discussed in (\ref{zigzag-3}), its Alexander polynomial can be written as follows
\begin{equation}
    \Delta_{{3}_1}(x) = x^{-1}-1+x=x^{-1}\big(1-x(1-x)\big) = 1 + \frac{(1-x)^2}{x}.   \label{Delta-31}
\end{equation}

First, we consider the ``homological approach''. It is based on the middle expression above, which is (\ref{Delta-f}) with $p=1$ and $f(x)=x$. From the inverse binomial theorem we find that (\ref{inverse-Delta}) reads
\be
\frac{1}{\Delta_{3_1}^{N-1}(x)}  = x^{N-1} \sum_{m=0}^\infty {N+m-2 \choose m} x^m (1-x)^m. \label{inverse-Delta-31}
\ee

Let us reconstruct now colored HOMFLY-PT polynomials. Following our prescription (\ref{Pr-q-deform}), the terms ${N+m-2 \choose m}  (1-x)^m$ are deformed into ${r \brack m} (aq^{-1};q)_m$, and we have $x^{N-1}=\frac{a^r}{q^r}$ and $x^m=q^{rm}$, so that colored HOMFLY-PT polynomials are expected to take form (note that the range of summation is in fact limited by $r$)
\be
P_r^{3_1}(a,q) = \frac{a^r}{q^r} \sum_{m=0}^r {r \brack m} q^{rm}q^{\alpha m^2 + \beta m} a^{\gamma m} (aq^{-1};q)_m,   \label{P31-a-b-c}
\ee
with additional potential deformation specified by $\alpha, \beta$ and $\gamma$ that still need to be determined. These parameters can be fixed in various ways. First, one can compare the above formula with known expressions for small colors. Uncolored HOMFLY-PT polynomial $P_1^{3_1} (a,q)$ follows from the table \ref{tab-P}, and $r=2$ polynomial is also known 
\be 
P_2^{3_1} (a,q) 
= a^2 q^{-2}   + a^2q(1+q) (1-aq^{-1})  + a^2 q^4(1-aq^{-1}) (1-a). 
\ee 
Comparing (\ref{P31-a-b-c}) for $r=2$ with these first two colored polynomials suffices to fix $\alpha, \beta$ and $\gamma$.

Alternatively, to fix (\ref{P31-a-b-c}), one can expand it in $\hbar$  and compare with higher order coefficients $R_k$ in (\ref{mmr-intro}), if they would be known independently. In general, because the number of parameters to be fixed in (\ref{Pr-q-deform}) is finite, it is sufficient to consider finite number of $R_k$ coefficients. For completeness we find that
\begin{equation}
    R_1^{3_1}(x,N)=-\frac{(N-1) (x-1) \left(N ((x-1) x+2)-2 \left(x^3+1\right)\right)}{2 x^2},
    \label{eq:R1}
\end{equation}
which for $N=2$ reduces to the result given in \cite{Gukov:2019mnk}.

The above comparisons yield $\alpha=\gamma=0$ and $\beta=1$, and lead to the final result 
\be 
P_r^{3_1}(a,q)  = \frac{a^r}{q^r} \sum_{m=0}^r {r \brack m} q^{m(r+1)} (aq^{-1};q)_m.   \label{Pr-31-h}
\ee 
Analogous computation, however including $t$-dependence, leads to the following form of colored superpolynomials
\be 
P_r^{3_1}(a,q,t)  = \frac{a^r}{q^r} \sum_{m=0}^r {r \brack m} q^{m(r+1)} t^{2m} (-aq^{-1}t;q)_m.   \label{Pr-31-h-t}
\ee 
These results are in agreement with colored superpolynomials found by other means in \cite{Fuji:2012pi}.

Furthermore, we note that the above analysis yields various information about the corresponding quiver. The expression (\ref{P31-a-b-c}), after $q$-binomial expansion of $(aq^{-1};q)_m$, can be written as
\begin{align}
\begin{split}
& P_r^{3_1}(a,q)  = \frac{a^r}{q^r} \sum_{0\le j \le m}^\infty 
\frac{(q;q)_r}{(q;q)_{r-m}(q;q)_{m-j}(q;q)_j}\, q^{rm+ \alpha m^2 + \beta m} a^{\gamma m}.
(-a)^j   q^{\frac{j}{2}(j-3)} = \\
& \quad =   \sum_{d_1+d_2+d_3=r} 
\frac{(q;q)_r(-1)^{d_3} a^{r+\gamma (d_2+d_3)+d_3}}{(q;q)_{d_1}(q;q)_{d_2}(q;q)_{d_3}}\, q^{(d_1+d_2+d_3)(d_2+d_3)+\frac{d_3}{2}(d_3-3)+ \alpha (d_2+d_3)^2 + \beta (d_2+d_3)-r} ,
\end{split}  \label{P31r-quiver}
\end{align}
where summations in the second line are rewritten in terms of $d_1 = r-m, d_2 = m-j$ and $d_3 = j$, so that the second line takes form of a quiver generating series \cite{Kucharski:2017ogk}. Recall that the underlying quiver is determined only by powers of $q$ quadratic in $d_i$. Note that the dependence on $d_1$ arises only from the coefficient $rm=(d_1+d_2+d_3)(d_2+d_3)$, which thus determines the first row (and column) of the quiver matrix as $C_{1,1}=0$ and $C_{1,2}=C_{1,3} = 1$, in agreement with \cite{Kucharski:2017poe}. This shows that even without fixing parameters $\alpha, \beta$ and $\gamma$, we can deduce at least a partial information about the corresponding quiver. Furthermore, $\alpha$ is the only parameter that determines remaining quadratic terms, and it can be determined already from the first correction (\ref{eq:R1}) in the Melvin-Morton-Rozansky expansion (\ref{mmr-intro}). Thus this first correction is sufficient to specify the whole quiver (even though additional parameters $\beta$ and $\gamma$ need to be determined by further corrections $R_k$, or by other means). Fixing $\alpha=0$, as explained above, from quadratic terms in the power of $q$ in (\ref{P31r-quiver}) we find that the whole quiver matrix takes form
\be
C = \left[\begin{array}{ccc}
0 & 1 & 1 \\
1 & 2 & 2 \\
1 & 2 & 3
\end{array}\right]    \label{C31}
\ee
in agreement with the results in \cite{Kucharski:2017ogk}. Note that this quiver consists of 3 nodes, which are in one-to-one correspondence with generators of HOMFLY-PT homology, and diagonal elements of the above matrix are $t$-degrees of these generators. The fact that we immediately obtain this particular quiver is the feature of the particular form of (\ref{Pr-31-h}).

Let us now discuss the second, ``cyclotomic'' form of expansion. In this case we consider Alexander polynomial written as in the expression on the right in (\ref{Delta-31}), which corresponds to $g(X)=-X$ in (\ref{Delta-d}). In this case the inverse binomial expansion (\ref{inverse-Delta2}) yields
\be
\frac{1}{\Delta_{3_1}^{N-1}(x)}  =  \sum_{m=0}^\infty (-1)^m {N+m-2 \choose m} \frac{ (1-x)^{2m}}{x^m} =  \sum_{m=0}^\infty (-1)^m (N-1)_m \frac{ (1-x)^{2m}}{m! x^m}. \label{inverse-Delta-31-c}
\ee
To determine colored HOMFLY-PT polynomial, we deform the Pochhammer $(N-1)_m$ to $(aq^{-1};q)_m$, and taking into account (\ref{deform-cyclo}) we get
\be
P_r^{3_1}(a,q)  =  \sum_{m=0}^r (-1)^m {r \brack m} q^{-rm} q^{\alpha m^2 + \beta m} a^{\gamma m} (aq^{-1};q)_m (aq^r;q)_m.   \label{Pr-31-c-abc}
\ee
Fixing $\alpha,\beta$ and $\gamma$ as above, by comparison to the first and the second colored polynomial, we find $\alpha=\beta=1/2$ and $\gamma=0$, so that finally
\be
P_r^{3_1}(a,q)  =  \sum_{m=0}^r (-1)^m {r \brack m} q^{-rm + (m^2 + m)/2}  (aq^{-1};q)_m (aq^r;q)_m  . \label{Pr-31-c}
\ee
Analogous computation, however including dependence of $t$, leads to colored superpolynomial
\be
P_r^{3_1}(a,q,t)  =  (-t)^{-r} \sum_{m=0}^r (-1)^m {r \brack m} q^{-rm + (m^2 + m)/2}  (-aq^{-1}t;q)_m (-aq^rt^3;q)_m  . \label{Pr-31-c-t}
\ee
This expression is of cyclotomic form (\ref{cyclo-t}), which is different than (\ref{Pr-31-h-t}), but of course both these results yield the same $P_r^{3_1}(a,q,t)$. However, for (\ref{Pr-31-c}) straightforward manipulations lead to a quiver of size 5 (see also \cite{Kucharski:2017ogk}), which has two extra generators compared to (\ref{C31}). Therefore we obtain the cyclotomic form at the expense of loosing a nice correspondence between quiver nodes and homology generators.

Finally, we stress that for trefoil knot, the ``homological'' reconstruction scheme enables to determine the $a$-deformed invariant $F_{K}(x,a,q)$, as already advertised in (\ref{FK-xaq}). This invariant arises from similar deformations of the expression (\ref{inverse-Delta-31}) as above. We simply express the result in terms of $x$ (i.e. without substituting $x=q^r$), and also leave an infinite range of summation over $m$. We again stress that the entire dependence on $x$ arises already from the expression (\ref{inverse-Delta-31}), while the subsequent deformation introduces only $x$-independent corrections. Identifying $F_{3_1}(x,a,q)$ in this way is possible, because $x$ arises only in positive powers in (\ref{inverse-Delta-31}), so that we obtain a well defined series in powers of $x$. Removing the overall normalization factor $x^{N-1}$ we obtain then the same result as in \cite{Ekholm:2020lqy}
\be
F_{3_1}(x,a,q) = \sum_{m=0}^{\infty} x^m q^m \frac{(x;q^{-1})_m (aq^{-1};q)_m}{(q;q)_m}.
\ee
Note that the second approach, which leads to (\ref{inverse-Delta-31-c}), is not suitable in this case and does not lead to $F_{3_1}(x,a,q)$ invariant, because of the presence of both positive and negative powers of $x$ in the summand of (\ref{inverse-Delta-31-c}).


\subsection{Figure-eight knot, $4_1$}    \label{ssec-41}

\begin{figure}[htp]
    \centering
    \includegraphics[width=0.3\textwidth]{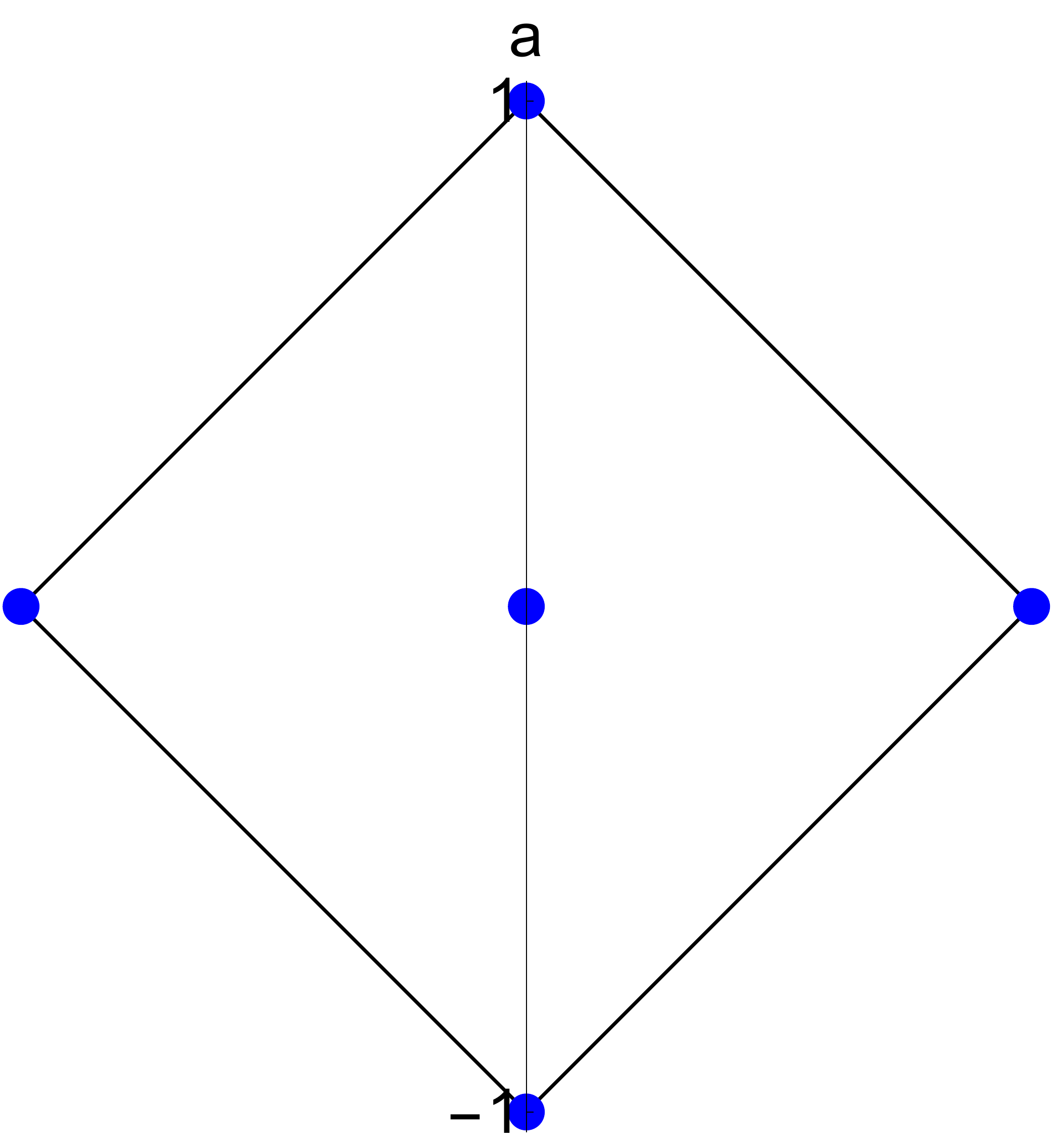}
    \caption{Homological diagram for $4_1$ knot.}\label{fig-41}
\end{figure}

Figure-eight knot is a prototype example of how to deal with pieces of Alexander polynomial corresponding to diamonds. The homological diagram for $4_1$ knot, shown in fig. \ref{fig-41}, consists of one diamond and a zig-zag of minimal length 1, and for the latter reason the ``homological'' and ``cyclotomic'' expansions essentially overlap.. The corresponding superpolynomial is given in table \ref{tab-P}. It is not hard to deduce its structure from Alexander polynomial, which we write accordingly
\begin{equation}
    \Delta_{{4}_1}(x) =- x^{-1} + 3 - x 
    =1-\frac{(1-x)^2}{x},
\end{equation}
which is automatically in the form relevant for cyclotomic expansion (\ref{Delta-d}), with $g(X)=X$. 

The formulae (\ref{inverse-Delta}) and (\ref{inverse-Delta2}) both yield
\begin{equation}  \label{eq:MMR0fig8}
 \frac{1}{\Delta_{{4}_1}(x)^{N-1}}
  = \sum_{m=0}^\infty  {N+m-2 \choose m} \frac{(1-x)^{2m}}{x^m} 
  = \sum_{m=0}^\infty  (N-1)_m \frac{(1-x)^{2m}}{m!x^m}. 
\end{equation}
Note that it differs from the result for trefoil (\ref{inverse-Delta-31-c}) only by a sign $(-1)^m$. Therefore its deformation is analogous; we promote the Pochhammer $(N-1)_m$ to $(aq^{-1};q)_m$, and taking into account (\ref{deform-cyclo}) we get
\be 
P_r(a,q) = \sum_{m=0}^r  {r \brack m}  a^{\alpha m} q^{\beta m^2 + \gamma m} q^{-rm} (aq^{-1};q)_m (aq^r;q)_m,        \label{Pr-41-a-b-s}
\ee 
where $\alpha,\beta$ and $\gamma$ encode potential further deformations yet to be specified. These parameters can be fixed by comparison with the uncolored HOMFLY-PT polynomial following from table \ref{tab-P}, and the polynomial in the second symmetric representation
\be 
\begin{split}
P_2^{4_1} (a,q) &= 1 +  a^{-1}q^{-1} (1+q)(1-aq^{-1}) (1-aq^2)  + \\
& \qquad + a^{-2}q^{-2} (1-aq^{-1}) (1-aq^2) (1-a) (1-aq^3).
\end{split}
\ee 
Alternatively, one can compare subleading terms in the Melvin-Morton-Rozansky expansion of (\ref{Pr-41-a-b-s}) with the polynomials $R_k(x,N)$ in (\ref{mmr-intro}). For example, the first subleading correction in this expansion takes form by
\begin{equation}
    R_1^{4_1}(x,N)=-\frac{(N-2) (N-1) \big((x-3) x+1\big)  (x^2-1)}{2 x^2}.
\end{equation}
In particular, from the first two colored polynomials we fix $\alpha=-1, \beta=0$ and$\gamma=1$, so that
\be 
P_r^{4_1} (a,q) = \sum_{m=0}^r {r \brack m} a^{-m} q^{-rm+ m} (aq^{-1};q)_m (aq^r;q)_m.
\ee 
Including $t$-dependence in the above computation, we analogously find the formula for colored superpolynomials
\be 
P_r^{4_1} (a,q,t) = \sum_{m=0}^r {r \brack m} a^{-m} q^{-rm+ m}t^{-2m} (-aq^{-1}t;q)_m (-aq^rt^3;q)_m.
\ee 
These expressions are in agreement with the results in \cite{Kucharski:2017ogk}. Note that, as expected, we get a cyclotomic expression (\ref{cyclo-t}).

We can also relate the above analysis to the knots-quivers correspondence. Because in this case  ``cyclotomic'' and ``homological'' expansions overlap, we now find a quiver whose size agrees with the number of HOMFLY-PT generators. We simply rewrite  (\ref{Pr-41-a-b-s}) using the identity \eqref{identity-qpoch} 
\be
\begin{split} 
P_r (a,q) & = \, \sum_{0\le j \le k}^\infty \sum_{l=0}^{k-j} \sum_{m=0}^j (-1)^{j+l+m}\frac{(q;q)_r}{(q;q)_{r-k}(q;q)_{k-j-l} (q;q)_l (q;q)_{j-m}(q;q)_m}  \times \\ 
& \qquad \qquad \times
 q^{\frac12(l^2+m^2-3(l+m))} q^{\beta k^2 + \gamma k}
\, q^{jl + \frac{j}{2}(j-1)+rj-rk } a^{l+m+j+\alpha k} .
\end{split}
\ee  
Then setting $d_1=r-k, d_2 = k-j-l, d_3=j-m, d_4=l$ and $d_5 =m$ brings this expression to the form of a quiver generating series. As before, the first row (column) of the quiver matrix is determined even before fixing parameters $\alpha,\beta$ and $\gamma$: this first row arises from quadratic powers of $q$ of the form $r(j-k)=(d_1+\ldots +d_5)(-d_2-d_4)$, which implies that $C_{11} = C_{13}=C_{15}=0, C_{12}=C_{14}=-1$. Fixing parameters $\alpha,\beta$ and $\gamma$ gives rise to the full quiver then, in agreement with \cite{Kucharski:2017ogk}.


\subsection{$5_1$ knot}

The analysis for $5_1$ knot is analogous to the trefoil. However, in this case extra summations in expressions for colored polynomials arise, so we also discuss this example in detail, albeit focusing only on the ``homological'' expansion. Homological diagram for $5_1$ knot consists of a zig-zag of length $5$, so its Alexander polynomial takes form (\ref{zigzag-p}) with $p=2$
\be
\Delta_{5_1}(x) = \frac{1}{x^2} \big(1-x (1-x) (1+x^2)\big),
\ee
so that (\ref{fx}) takes form $f(x) = x(1+x^2)$, and then (\ref{inverse-Delta}) reads
\be
\begin{split}
\frac{1}{\Delta_{5_1}(x)^{N-1}} &= x^{2(N-1)} \sum_{m=0}^{\infty} {N+m-2 \choose m} x^m (1-x)^m (1+x^2)^m = \\
& = x^{2(N-1)} \sum_{0\leq k_2 \leq k_1}^{\infty} {N+k_1-2 \choose k_1} {k_1 \choose k_2} x^{k_1+2k_2} (1-x)^{k_1}.    \label{inverse-Delta-51}
\end{split}
\ee

For colored HOMFLY-PT polynomials, our quantization rules now yield
\be
P_r^{5_1} (a,q) 
=  \frac{a^{2r}}{q^{2r}} \sum_{0 \le k_2 \le k_1}^r {r \brack k_1}{k_1\brack k_2} q^{r(k_1+2k_2)+\alpha(k_1,k_2)} a^{\beta_1 k_1 + \beta_2 k_2} (aq^{-1};q)_{k_1},
\ee
where $\alpha(k_1,k_2)$ is a quadratic polynomial in $k_1$ and $k_2$, which together with parameters $\beta_1$ and $\beta_2$ still needs to be fixed. By comparing with HOMFLY-PT polynomials for $r=1$ (in table \ref{tab-P}) and for $r=2$
\be
\begin{split}
P_2^{5_1} (a,q) & = a^2 q^{-2} \big( 1 + (1+q) (1-aq^{-1}) (q^3 + q^{7}) + \\ 
& \qquad \qquad 
+ (1-a q^{-1}) (1-a) (q^{6} + (1+q)q^{9} + q^{12})\big)
\end{split}
\ee
we find $\alpha(k_1,k_2) = k_1+k_2 - k_1k_2$ and $\beta_1=\beta_2=0$, so that
\be
P_r^{5_1} (a,q) =  \frac{a^{2r}}{q^{2r}} \sum_{0 \le k_2 \le k_1}^r {r \brack k_1}{k_1\brack k_2} q^{r(k_1+2k_2)+k_1+k_2 - k_1k_2} (aq^{-1};q)_{k_1}.
\ee
Including dependence on $t$ in the computation, we analogously reconstruct colored superpolynomials
\be
P_r^{5_1} (a,q,t) =  \frac{a^{2r}}{q^{2r}} \sum_{0 \le k_2 \le k_1}^r {r \brack k_1}{k_1\brack k_2} q^{r(k_1+2k_2)+k_1+k_2 - k_1k_2} t^{2(k_1+k_2)} (-aq^{-1}t;q)_{k_1}.
\ee
This expression is not cyclotomic (to find such form one should follow the ``cyclotomic'' expansion), however it can be immediately rewritten in the quiver form, for a quiver with 5 nodes, which are in one-to-one correspondence with HOMFLY-PT generators. 

Furthermore, following (\ref{FK-xaq}), and similarly as in the case of trefoil, the analogous transformations enable us to reconstruct the invariant $F_{5_1}(x,a,q)$. The whole dependence on $x$ follows simply from (\ref{inverse-Delta-51}). Removing normalization factor $x^{2(N-1)}$ we then reproduce (in agreement with \cite{Ekholm:2020lqy})
\be
F_{5_1}(x,a,q) = \sum_{0 \le k_2 \le k_1}^r {k_1\brack k_2} x^{k_1+2k_2} q^{k_1+k_2 - k_1k_2} \frac{(x;q^{-1})_{k_1} (aq^{-1};q)_{k_1}}{(q;q)_{k_1}}.
\ee


\subsection{$(2,2p+1)$ torus knots}

Analogously to the $5_1$ case, we can reconstruct colored polynomials for $(2,2p+1)$ torus knots for any $p$. For brevity, we also focus only on the ``homological'' expansion. For a given $p$, Alexander polynomial takes form (\ref{zigzag-p}), and the inverse binomial theorem (\ref{inverse-Delta}) yields
\begin{equation}
\label{eq:torMMR0}
    \frac{1}{\Delta(x)^{N-1}}
= x^{p(N-1)}\sum_{0 \le k_p \le \ldots \le k_1}^\infty {N+k_1-2 \choose k_1} {k_1 \choose k_2}\cdots {k_{p-1} \choose k_p} x^{k_1+2k_2 + \ldots + 2k_p} (1-x)^{k_1} ,
\end{equation}
where multiple binomials arise from the expansion of $f(x)^m=(x\sum_{j=0}^{p-1} x^{2j})^m$. $q$-deformation and $a$-deformation of this expression yields the formula for colored HOMFLY-PT polynomials
\be 
P_r(a,q) 
= \frac{a^{pr}}{q^{pr}} \sum_{0 \le k_p \le ... \le k_1}^\infty {r \brack k_1}...{k_{p-1}\brack k_p} (aq^{-1};q)_{k_1}\,q^{r(k_1+2k_2+...+2k_p)} \, q^{\alpha(k_1,\ldots,k_p)} a^{\sum_i \beta_i k_1},   \label{Pr-22p1-a-b}
\ee 
where $\alpha(k_1,\ldots,k_p)$ is a quadratic polynomial in $k_i$ that needs to be specified together with parameters $\beta_i$, either by comparison with the first few colored polynomials, or with the first few coefficients $R_k$ in (\ref{mmr-intro}). Such a comparison yields the final result
\be 
P_r(a,q)
= \frac{a^{pr}}{q^{pr}} \sum_{0 \le k_p \le ... \le k_1}^r {r \brack k_1}\cdots{k_{p-1} \brack k_p} q^{(2r+1)(k_1+...+k_p) - rk_1-k_1k_2-...- k_{p-1}k_p} (aq^{-1};q)_{k_1}.
\ee 
Furthermore, including $t$-dependence in the above computation leads to the form of colored superpolynomials
\be 
P_r(a,q,t)
= \frac{a^{pr}}{q^{pr}} \sum_{0 \le k_p \le ... \le k_1}^r {r \brack k_1}\cdots{k_{p-1} \brack k_p} q^{(2r+1)(\sum_{i=1}^p k_i) - rk_1-\sum_{i=1}^{p-1}k_i k_{i+1}  }t^{2\sum_{i=1}^p k_i} (-aq^{-1}t;q)_{k_1}
\ee 
matching the results in \cite{Kucharski:2017ogk}. 

In this general case let us also identify the information about the corresponding quiver. Using (\ref{identity-qpoch}) we can rewrite (\ref{Pr-22p1-a-b}) in the form
\be 
\begin{split}
P_r(a,q) & = \frac{a^{pr}}{q^{pr}} \sum_{0 \le k_p \le ... \le k_1}^\infty \sum_{l_1=0}^{k_1-k_2} \ldots \sum_{l_p=0}^{k_p}   q^{r(k_1+2k_2+...+2k_p)} \, q^{\alpha(k_1,\ldots,k_p)} a^{\sum_i \beta_i k_1} \times  \\ 
& \quad \times \frac{(-a)^{l_1+...+l_p}q^{\frac12\sum_{i=1}^p(l_i^2-3l_i)}q^{k_1(l_2+...+l_p) -k_2l_2 ...-k_pl_p}}{(q;q)_{k_1-k_2-l_1}(q;q)_{l_1}...(q;q)_{l_{p_1}}(q;q)_{k_{p-1}-k_p-l_{p-1}}(q;q)_{l_{p-1}}(q;q)_{l_p}(q;q)_{k_p-l_p}}.
\end{split}
\ee 
Changing the variables as follows
$d_1=r-k_1, d_2= l_1,d_3=k_1-k_2-l_1,\ldots, d_{2p-2} = l_{p-1}, d_{2p-1}=k_{p-1}-k_p-l_{p-1}, d_{2p}=l_p, d_{2p+1} = k_p - l_p$, the denominators in the above expression are turned to $\prod_{i=1}^{2p+1} (q;q)_{d_i}$, and the whole expression takes form of a quiver generating function. The form of the quiver arises from quadratic powers of $q$, and in particular the first row arises from terms proportional to $r=\sum_i d_i$, which is the only source of $d_1$. There is one such term
\be 
\begin{split}
& r(k_1 + 2k_2 + ...+ 2k_p) = \\ 
& \quad = r(l_1+(k_1-k_2-l_1)) + 3r(l_2 + (k_2-k_3-l_2)) + \ldots + (2p-1)r (l_p + (k_p-l_p)) = \\
& \quad = r\big(d_2 + d_3 + 3d_3 + \ldots + (2p-1)(d_{2p} + d_{2p+1})\big).
\end{split}
\ee
It follows that $C_{11}=0, C_{12}=C_{13} =1, C_{14}=C_{15}=3, \ldots, C_{1,2p}=C_{1,2p+1} = 2p-1$, so that we can fix the first row (and column) of the quiver even without fixing $\alpha(k_1,\ldots,k_p)$ and $\beta_i$. Once these parameters are fixed, we can then read off the whole quiver matrix, which is consistent with \cite{Kucharski:2017ogk}.
 
Finally, for this whole family of torus knots $K=T^{2,2p+1}$ we can also reconstruct $F_K(x,a,q)$ invariants, following (\ref{FK-xaq}), and analogously to $3_1$ and $5_1$ case. The whole $x$-dependence follows from (\ref{eq:torMMR0}), and after appropriate deformation, and removing $x^{p(N-1)}$ prefactor, we obtain the same result as in \cite{Ekholm:2020lqy}
  \be
F_{T^{2,2p+1}} = \sum_{0 \le k_p \le ... \le k_1} {k_1 \brack k_2}\cdots{k_{p-1} \brack k_p} x^{k_1+2\sum_{i=2}^p k_i} q^{\sum_{i=1}^p k_i - \sum_{i=2}^p  k_{i-1}k_i} \frac{(x;q^{-1})_{k_1} (aq^{-1};q)_{k_1}}{(q;q)_{k_1}}.
\ee


\subsection{Twist knots $4_1, 6_1, 8_1,\ldots$}
     
\begin{figure}[htp]
    \centering
    \includegraphics[width=0.3\textwidth]{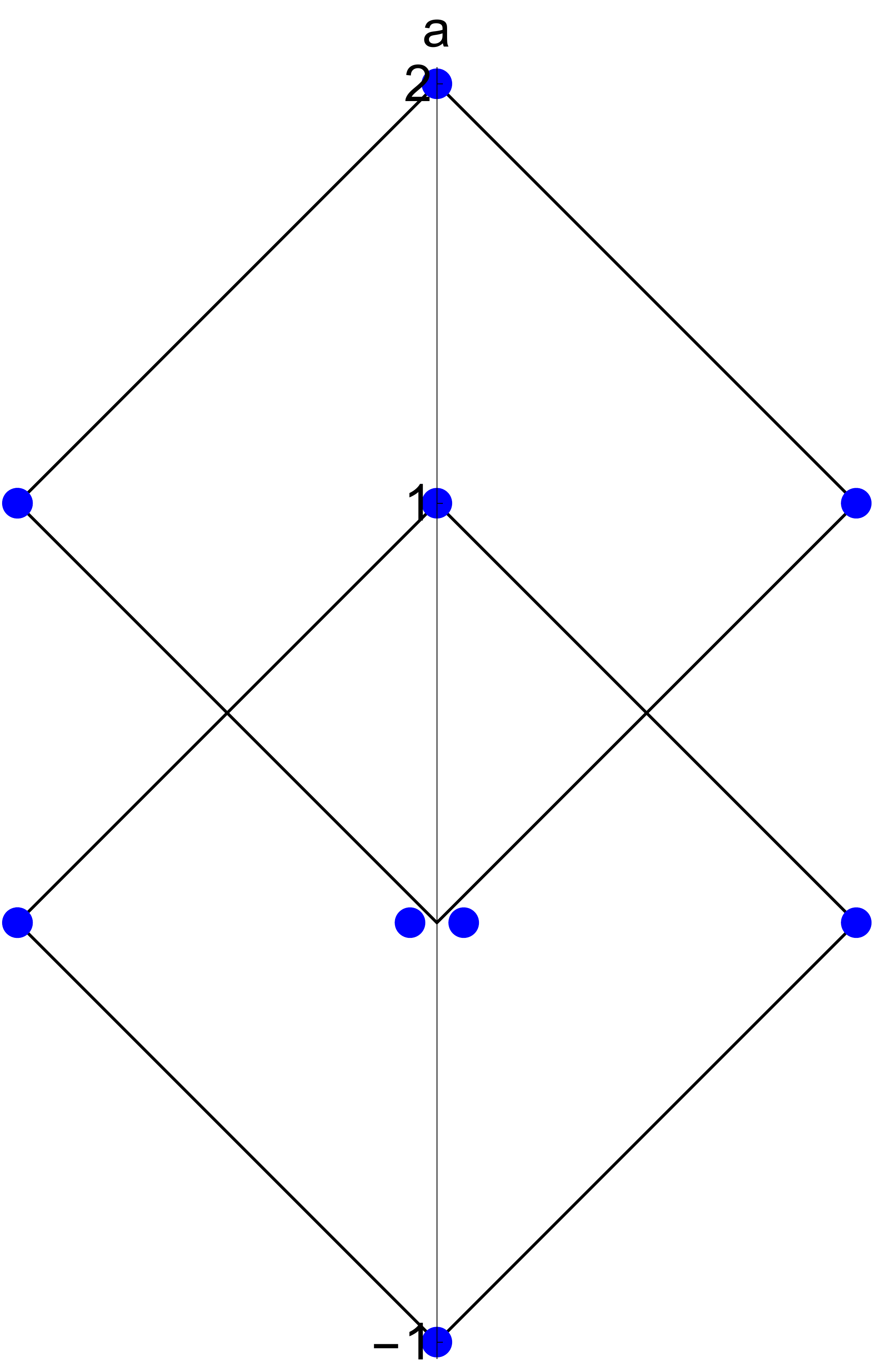}
    \caption{Homological diagram for $6_1$ knot.}\label{fig-61}
\end{figure}

We consider now a family of twist knots $(2p+2)_1$, for $p=1,2,3,\ldots$. The case $p=1$ is the figure-eight knot $4_1$, and the analysis from section \ref{ssec-41} generalizes to other values of $p$. For a fixed $p$, a homological diagram consists of $p$ diamonds displaced vertically and a zig-zag of minimal length 1, as seen in the example of $6_1$ knot in fig. \ref{fig-61}. For a given $p$, Alexander polynomial takes form
\begin{equation}
    \Delta(x) = 1 - p \frac{(1-x)^2}{x}.    \label{Alexander-twist-I}
\end{equation}
Because the zig-zag is of length 1, this form is relevant for both ``homological'' and ``cyclotomic'' expansion. In this case the inverse binomial expansion takes form
\begin{equation}
    \frac{1}{\Delta(x)^{N-1}} = \sum_{0 \le k_p \le \ldots \le k_1 }^\infty   {N+k_1-2 \choose k_1}
{k_1 \choose k_2} {k_2 \choose k_3}\ldots{k_{p-1} \choose k_p}  \frac{(1-x)^{2k_1}}{x^{k_1}}.   \label{eq:tw1MMR0}
\end{equation}

Following the quantization and deformation prescription we find that colored HOMFLY-PT polynomials take form
\be 
P_r(a,q)=  \sum_{0 \le k_p \le \ldots \le k_1 }^\infty{r \brack k_1}\cdots{k_{p-1}\brack k_p}\, (aq^{-1};q)_{k_1}\, (aq^r;q)_{k_1} q^{-rk_1}
q^{\alpha(k_1,\ldots,k_p)} a^{\sum_{i=1}^p \beta_i k_i},
\ee 
where $\alpha(k_1,\ldots,k_p)$ is a quadratic polynomial in $k_i$, and $\beta_i$ are parameters. Fixing them by comparing with several first colored polynomials yields the same result as in \cite{Fuji:2012pi,Nawata:2012pg}
\be 
\begin{split}
P_r(a,q)& =  \sum_{0 \le k_p \le \ldots \le k_1 }^\infty{r \brack k_1}\cdots{k_{p-1}\brack k_p}\, (aq^{-1};q)_{k_1} (aq^r;q)_{k_1} q^{-rk_1+k_1}a^{-k_1+k_2+\ldots+k_p} \times \\ 
& \qquad \times q^{(k_2^2+\ldots+k_p^2)-(k_2+k_3+\ldots+k_p)}.   \label{Pr-twist-I}
\end{split}
\ee 
The generalization to the colored superpolynomials yields 
\be 
\begin{split}
P_r(a,q,t)& =  \sum_{0 \le k_p \le \ldots \le k_1 }^\infty{r \brack k_1}\cdots{k_{p-1}\brack k_p}\, (-aq^{-1}t;q)_{k_1} (-aq^rt^3;q)_{k_1} q^{-rk_1+k_1}a^{-k_1+k_2+\ldots+k_p} \times \\ 
& \qquad \times q^{(k_2^2+\ldots+k_p^2)-(k_2+k_3+\ldots+k_p)}  t^{-2k_1+2(k_2+k_3+\dots + k_p)}.
\end{split}
\ee

Note that this result is of cyclotomic form (\ref{cyclo}), and we can also relate it to the corresponding quiver. The identity \eqref{identity-qpoch} yields
\be 
\begin{split}
 P_r^{(2p+2)_1}(a,q) &=  \sum_{0 \le k_p \le ... \le k_1 }^\infty    q^{-rk_1}q^{\alpha(k_1,\ldots,k_p)} a^{\sum_{i=1}^p \beta_i k_i} \frac{(q;q)_r}{(q;q)_{r-k_1}} \times
 \\ & \qquad \times
 \sum_{l_1=0}^{k_1-k_2} \cdots \sum_{l_p=0}^{k_p} (-aq^{r})^{l_1+...+l_p}q^{\frac12 \sum_{i=1}^p(l_i^2-l_i)}
 q^{k_1(l_2+...+l_p)-k_2l_p-...-k_pl_p} \times
 \\ & \qquad \times 
\sum_{m_1=0}^{l_1}\sum_{n_1=0}^{k_1-k_2-l_1}\cdots \sum_{m_p=0}^{l_p}\sum_{n_p=0}^{k_p-l_p}
\frac{(-a)^{m_1+n_1+...+m_p+n_p}q^{\frac12\sum_{i=1}^p (m_i^2+n_i^2-3m_i-3n_i)}}{(q;q)_{m_1}(q;q)_{n_1}\cdots(q;q)_{m_p}(q;q)_{n_p} } \times
\\ & \qquad \times 
\frac{q^{T_{F1}}}{(q;q)_{l_1-m_1}(q;q)_{k_1-k_2-l_1-n_1}\cdots (q;q)_{l_p-m_p}(q;q)_{k_p-l_p-n_p}}
\end{split}
\ee 
where 
\be 
\begin{split}
 T_{F1}&= (k_1(n_1+...+n_p) -k_2n_2-...-k_pn_p)  +  \\
 & +l_1(m_2+...+m_p) + l_2(n_1+m_3+...+m_p)+l_3(n_1+n_2+m_4+...+m_{p-1}) + \\
 & +l_4(n_1+n_2+n_3+m_5+...+m_{p-1})+ ...+ l_{p-1}(n_1+...+n_{p-1}+m_p).
\end{split}
\ee 
Changing the summation variables in the above formula as $d_1=r-k_1, d_2 = k_1-k_2-l_1-m_1,d_3=l_1-n_1,d_4=l_1,d_5=n_1$, and so on, leads to the form of the quiver generating function. As before, we can now identify the first row of the quiver matrix, which arises from quadratic powers of $q$ proportional to $r$. In this way we find $C_{12}=C_{14}=\ldots=C_{1,2p-2}=C_{1,2p} = -1$, and all other entries are zero. Furthermore, after fixing $\alpha(k_1,\ldots,k_p)$ and $\beta_i$ as in (\ref{Pr-twist-I}), we reconstruct the full quiver in agreement with \cite{Kucharski:2017ogk}.


\subsection{Twist knots $3_1, 5_2, 7_2,\ldots$}   \label{ssec-twist-II}
     
\begin{figure}[htp]
    \centering
    \includegraphics[width=0.3\textwidth]{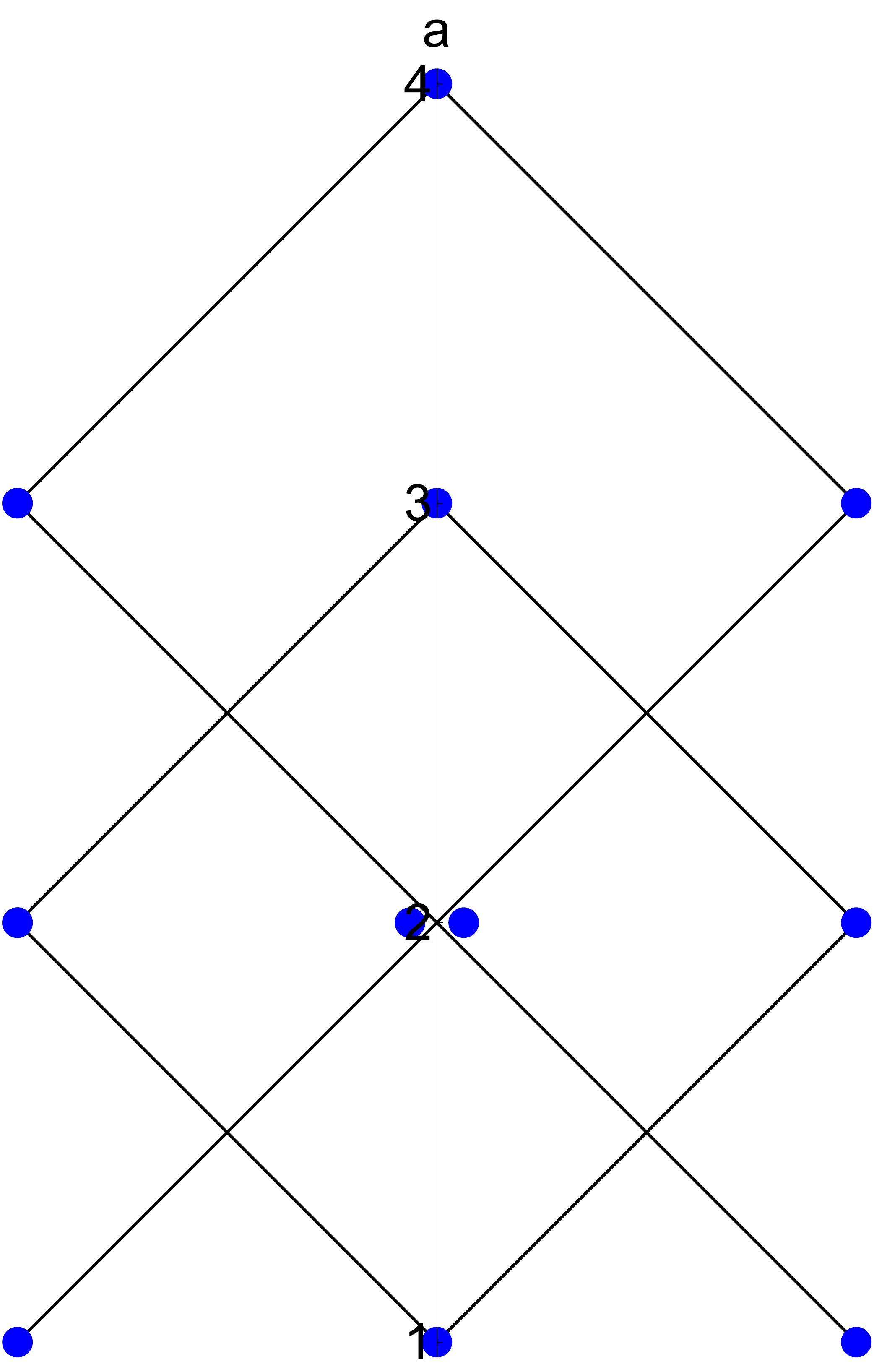}
    \caption{Homological diagram for $7_2$ knot.}\label{fig-72}
\end{figure}

The second family of twist knots consists of knots $3_1, 5_2, 7_2,\ldots$, which we label respectively by $p=1,2,3,\ldots$. Their homological diagrams consist of a zig-zag of length 3 and $(p-1)$ diamonds displaced vertically, and have $4p-1$ generators, as shown in the example in fig. \ref{fig-72}. Trefoil, the first knot in this family, is a little special, and its diagram consists of a zig-zag only. We write Alexander polynomial for knots in this family in the form relevant for ``cyclotomic'' expansion
\be
\Delta(x) = 1 + p \frac{(1-x)^2}{x}.
\ee
Note that it differs only by a sign from the result for the previous family of twist knots (\ref{Alexander-twist-I}). From the inverse binomial theorem we get
\be 
\frac{1}{\Delta^{N-1}(x)} =\sum_{0\le k_p\le \ldots\le k_1}^\infty (-1)^{k_1}{N+k_1-2 \choose k_1}
{k_1 \choose k_2} {k_2 \choose k_3}\cdots{k_{p-1} \choose k_p} \frac{ (1-x)^{2k_1} }{x^{k_1}}.
\ee 

As usual, appropriate deformation of this expression, including (\ref{deform-cyclo}), yields
\be 
P_r (a,q) 
= \sum_{0\le k_p\le \ldots \le k_1}^r 
(-1)^{k_1} {r \brack k_1}{k_1 \brack k_2} \cdots{k_{p-1} \brack k_p} (aq^{-1};q)_{k_1}
(aq^{r};q)_{k_1}  q^{-rk_1}q^{\alpha(k_1,\ldots,k_p)}   a^{\sum_{i} \beta_i k_i},
\ee
where $\alpha(k_1,\ldots,k_p)$ is a quadratic polynomial in summation variables $k_i$. By comparison with first colored polynomials one can fix the form of $\alpha(k_1,\ldots,k_p)$ and $\beta_i$, which leads to the results consistent with \cite{Nawata:2012pg}. We have already explicit formula for colored polynomials for $3_1$ knot in (\ref{Pr-31-c}). As another example, explicit formulae for $5_2$ knot and $9_2$ take form
\be 
P_r^{5_2}(a,q) = \sum_{0\le k_2\le k_1}^r (-1)^{k_1} {r \brack k_1}{k_1 \brack k_2} a^{k_2}
q^{-k_1r+ (k_1^2+k_1)/2 + (k_2^2-k_2) } (aq^{-1};q)_{k_1} (aq^r;q)_{k_1}
\ee
and  
\be 
\begin{split}
 P_r^{9_2} (a,q) 
& = \sum_{0 \le k_4 \le k_3 \le k_2 \le k_1}^\infty (-1)^{k_1} {r \brack k_1}{k_1 \brack k_2}{k_2 \brack k_3}{k_3 \brack k_4} (a q^{-1};q)_{k_1}(aq^{r};q)_{k_1} \times
\\ 
& \qquad \qquad \times q^{-k_1 r + (k_1^2+k_1)/2+ (k_2^2+k_3^2+k_4^2-k_2-k_3-k_4)}   a^{k_2+k_3+k_4}   .
\end{split}
\ee
For arbitrary twist knot in this family, i.e. for trefoil corresponding to $p=1$, and for  $(2p+1)_2$ knot for $p>1$, we find
\be 
\begin{split}
P_r (a,q) & = \sum_{0\le k_p\le \ldots \le k_1}^r 
(-1)^{k_1} {r \brack k_1}{k_1 \brack k_2} \cdots {k_{p-1} \brack k_p} (aq^{-1};q)_{k_1}
(aq^{r};q)_{k_1}  \times \\
& \qquad \qquad \times q^{-rk_1+(k_1^2+k_1)/2 + \sum_{i=2}^p(k_i^2 - k_i  )}  a^{ \sum_{i=2}^p k_i }.   \label{Pr-twistII}
\end{split}
\ee
Furthermore, we find that the $t$-dependent colored superpolynomials for this class of knots read 
\be 
\begin{split}
P_r (a,q,t) & = \sum_{0\le k_p\le \ldots \le k_1}^r 
(-1)^{r+k_1} {r \brack k_1}{k_1 \brack k_2} \cdots {k_{p-1} \brack k_p} (-aq^{-1}t;q)_{k_1}
(-aq^{r}t^3;q)_{k_1}  \times \\
& \qquad \qquad \times q^{-rk_1+(k_1^2+k_1)/2 + \sum_{i=2}^p(k_i^2 - k_i  )}  a^{ \sum_{i=2}^p k_i } t^{ 2\sum_{i=2}^p k_i - r }.   \label{Pr-twistII-t}
\end{split}
\ee
These results are of cyclotomic form. 

For this family of twist knots, using the identity (\ref{identity-qpoch}), one can also find corresponding quivers, and deduce their partial structure even before determining all deformation parameters. However, starting from the expression (\ref{Pr-twistII}), we would obtain a quiver whose size is larger than the number of homology generators. To obtain a quiver of appropriate size we should find an expression for colored polynomials following the ``homological'' expansion, which we skip for brevity.


\subsection{$6_2$ knot}

\begin{figure}[htp]
    \centering
    \includegraphics[width=0.45\textwidth]{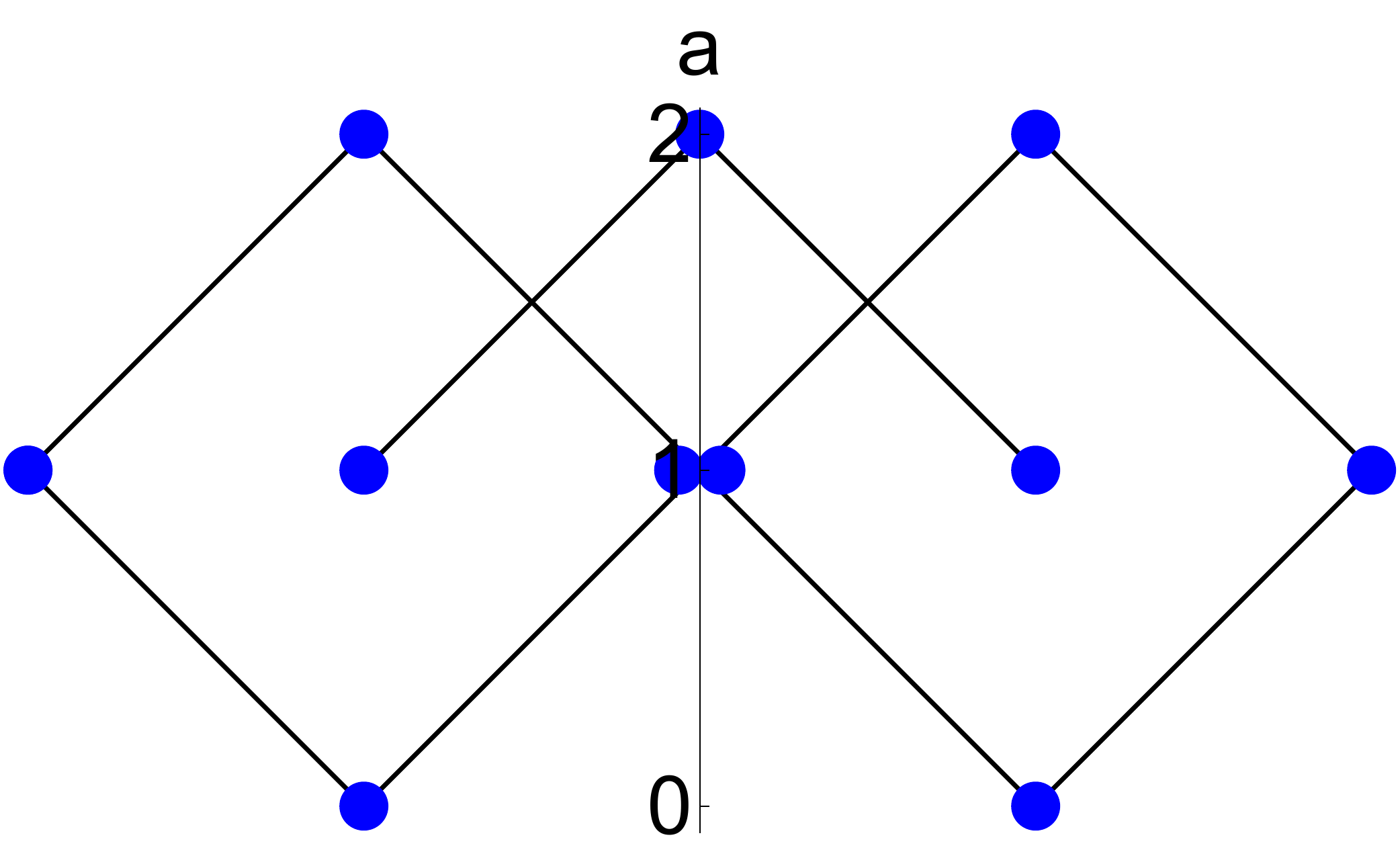}
    \caption{Homological diagram for $6_2$ knot.}\label{fig-62}
\end{figure}

We consider now $6_2$ knot. Interestingly, diamonds in its homological diagram are displaced horizontally as shown in fig. \ref{fig-62}, which results in a bit more involved analysis than in other examples. Colored superpolynomials for this knot have been determined in \cite{Nawata:2015wya}, so we will illustrate how those results can be reproduced from our perspective. First, we rewrite Alexander polynomial in the form 
\be
\begin{split} 
\Delta_{6_2} (x) & = -x^2-x^{-2} + 3x + 3x^{-1} -3 = \\
& = \underbrace{(x+x-1-x^2)}_{\textrm{diamond 1}} + \underbrace{(x^{-1}+x^{-1}-1-x^{-2})}_{\textrm{diamond 2}} + \underbrace{(-1+x+x^{-2})}_{\textrm{zig-zag of length 3}} = \\
& = 1-\frac{(1-x)^2}{x} - \frac{(1-x)^4}{x^2}.    \label{Alexander-62}
\end{split}
\ee 
In the middle line we indicated explicitly how various monomials are associated to a zig-zag and two diamonds in the homology diagram. In the third line we rewrite Alexander polynomial in the cyclotomic form, so that the inverse binomial expansion (\ref{inverse-Delta2}), with $g(X)=X(1+X)$ and $X=\frac{(1-x)^2}{x}$, yields 
\be
\frac{1}{\Delta(x)^{N-1}} = \sum_{m=0}^{\infty} {N+m-2 \choose m}  X^m (1-X)^m=  \sum_{k_1=0}^{\infty} c_{k_1} X^{k_1}. \label{inverse-Delta-62}
\ee
The coefficients $c_{k_1}$ arise from assembling fixed powers of $X$ and take form
\be
\begin{split}
c_{k_1} & = \sum_{l=0}^{\lfloor{k_1/2}\rfloor}{N+k_1-l-2\choose k_1-l}{k_1-l\choose l} = \\
& = \sum_{0\le k_3 \le k_2}^{k_1} (-1)^{k_2+k_3}{N+k_2-2\choose k_2}{k_2 \choose k_3}{N+k_1-k_2+k_3-2 \choose k_1-k_2}    \label{ck1-62}
\end{split}
\ee
where we used a non-trivial identity to get the expression in the second line.

We deform now the summand in (\ref{inverse-Delta-62}) as follows
\be 
\begin{split} 
c_{k_1}X^{k_1} &=(-1)^{k_2+k_3}  {N+k_2-2 \choose k_2}{k_2\choose k_3}{N+k_1-k_2+k_3-2 \choose k_1-k_2} \frac{(1-x)^{2k_1}}{x^{k_1}} \ \rightsquigarrow   \\ 
& \ \rightsquigarrow\   (-1)^{k_2+k_3} 
\frac{(q,q)_r}{(q,q)_{r-k_1}}
\frac{(q^{(N-1)};q)_{k_2}}{(q;q)_{k_2-k_3}(q;q)_{k_3}}\frac{(q^{(N+k_3-1)};q)_{k_1-k_2}}{(q;q)_{k_1-k_2}}\frac{(q^{N+r};q)_{k_1}}{ q^{rk_1} } = \\ 
&  \qquad= (-1)^{k_2+k_3} 
{r \brack k_1}{k_1\brack k_2}{k_2\brack k_3} (a q^{-1};q)_{k_2}(aq^{k_3-1};q)_{k_1-k_2}(aq^{r};q)_{k_1}  q^{-rk_1} ,
\end{split}     \label{q-deform-62}
\ee 
which then yields
\be 
\label{slN62a}
\begin{split} 
P_r (a,q) 
& = \sum_{0\le k_3\le k_2\le k_1}^r (-1)^{k_2+k_3} {r \brack k_1}{k_1\brack k_2}{k_2\brack k_3} 
q^{-rk_1 + \alpha(k_1,k_2,k_3)} a^{ \sum_{i=1}^3\beta_i k_i} \times
\\ & \qquad \qquad \times 
(aq^{-1};q)_{k_2} (a q^{k_3-1};q)_{k_1-k_2} (aq^{r};q)_{k_1},
\end{split}
\ee 
where $\alpha(k_1,k_2,k_3)$ is a quadratic polynomial. Its form, as well as parameters $\beta_i$, can be fixed by comparison with $t=-1$ specialization of the uncolored superpolynomial in table \ref{tab-P}, which can also be rewritten as follows (it is instructive to identify a zig-zag and diamonds from fig. \ref{fig-62} in this form)
\be
 P_\square (a,q) = -t^{-1} + (q^{-1}t^{-2}+t^{-1}+q) (1+aq^{-1}t) (1+aqt^3) ,
 \ee
and $S^2$-colored polynomial
\be
\begin{split} 
P_2  (a,q) = & \,
1 + (1+q)(q^{-2} + 1 - q^{-1}) (1-a/q)(1-aq^2) + \\ 
&  + q^{-4} (1-aq^{-1}) (1-a) (1-aq^2) (1-aq^3)  +\\ 
& + (q^{-1} - (1+q) + q^2) (1-aq^{-1}) (1-a) (1-aq^2) (1-aq^3) + \\
&  - q^{-2} (1+q)(1-aq^{-1}) (1-aq^{-1}) (1-aq^2) (1-aq^3) + \\ 
&    + q^{-1} (1+q)(1-aq^{-1}) (1-a) (1-aq^2) (1-aq^3). 
\end{split} 
\ee 
In particular, the terms in the last two lines above, independently fix the structure of $q$-Pochhammers in the second line of (\ref{slN62a}) and motivate the identity that we used in (\ref{ck1-62}). Note that more complicated structure of $q$-Pochhammers in the second line of (\ref{slN62a}) than in other examples is a consequence of horizontal displacement of two diamonds in the homological diagram. Having fixed $\alpha(k_1,k_2,k_3)$ and $\beta_i$, we find the final result
\be 
\begin{split} 
\label{NOHom}
P_r (a,q) &=\sum_{0\le k_3\le k_2\le k_1}^r  {r \brack k_1}{k_1\brack k_2}{k_2\brack k_3}  (a q^{-1};q)_{k_2}(aq^{k_3-1};q^2)_{k_1-k_2}(aq^{r};q)_{k_1}\times \\ 
& \qquad \qquad \times (-1)^{k_2+k_3}  q^{\frac12 k_3(k_3+1) - \frac12 k_2(k_2-1) + k_1k_2-rk_1},
\end{split}
\ee 
which is in agreement with the expression in \cite{Nawata:2015wya}, and proves that this expression is consistent with the Melvin-Morton-Rozansky theorem. It is also not difficult to include the $t$-dependent refinement. An analogous computation as above yields
\be
\begin{split}
\label{NOHOM-t}
P_r (a,q,t) &=  \sum_{0\le k_3\le k_2\le k_1}^r  {r \brack k_1}{k_1\brack k_2}{k_2\brack k_3}  (-a q^{-1}t;q)_{k_2}(-aq^{k_3-1}t;q^2)_{k_1-k_2}(-aq^{r}t^3;q)_{k_1} \times
\\ & \qquad \times
(-1)^{r-k_1} q^{\frac12 k_3(k_3+1)-\frac12 k_2(k_2-1) +k_1k_2-rk_1}  t^{k_2+k_2-k_1-r}.
\end{split}
\ee 
As expected, these final expressions have the cyclotomic form.


\subsection{$6_3$ knot}

\begin{figure}[htp]
    \centering
    \includegraphics[width=0.45\textwidth]{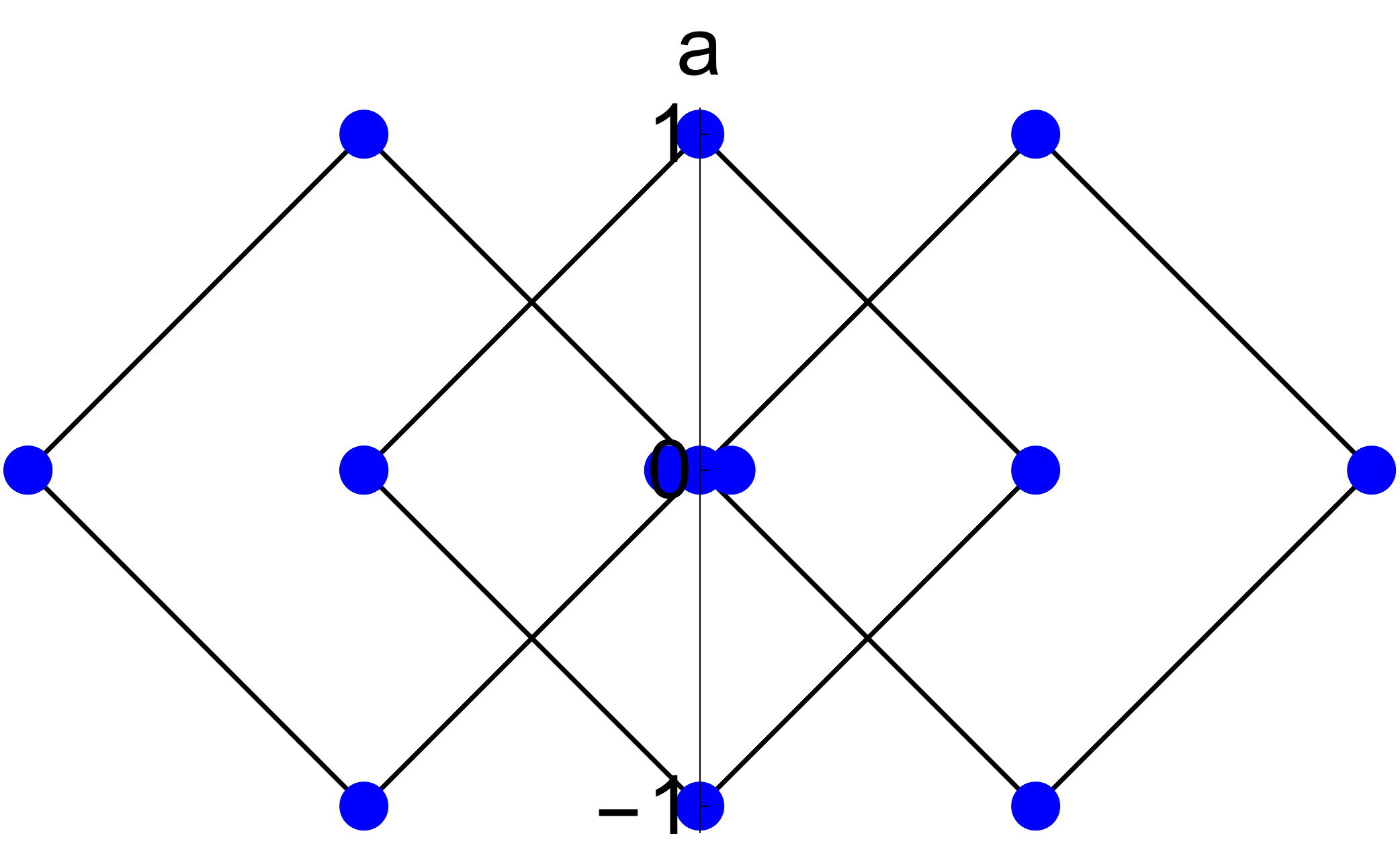}
    \caption{Homological diagram for $6_3$ knot.}\label{fig-63}
\end{figure}

For completeness we also discuss the knot $6_3$, whose analysis is very similar to the previous $6_2$ example. Its homological diagram consists of three diamonds and a zig-zag made of one dot, see fig. \ref{fig-63}. Its Alexander polynomial can be rewritten as follows
\be
\begin{split}
\Delta_{6_3}(x) &=  x^{-2} - 3x^{-1} +5   - 3x +x^2  = \\
& = \underbrace{1}_{\textrm{trivial zig-zag}} + \underbrace{(x^{-2}-2x^{-1}+1)}_{\textrm{diamond 1}} + \underbrace{(x^2-2x+1)}_{\textrm{diamond 2}} + \underbrace{(-x^{-1}+2-x)}_{\textrm{diamond 3}} = \\
& = 1 + \frac{(1-x)^2}{x} + \frac{(1-x)^4}{x^2}.
\end{split}
\ee
In the middle line we identify elements of the homological diagram. Note that the third line differs only in signs from $6_2$ example (\ref{Alexander-62}), which is a consequence of replacing a zig-zag of length 3 (for $6_2$) by a diamond and a zig-zag of length 1 (for $6_3$). Therefore we can conduct the inverse binomial expansion analogously as in (\ref{inverse-Delta-62}), also using the identity (\ref{ck1-62}), just taking care of the signs 
\be 
\frac{1}{\Delta_{6_3}^{N-1}(x)}
= \sum_{0\le k_3\le k_2\le k_1}^\infty 
(-1)^{k_1+k_3}
{N+k_2-2 \choose k_2}{k_2\choose k_3}{N+k_1-k_2+k_3-2 \choose k_1-k_2} \frac{(1-x)^{2k_1}}{x^{k_1}}.
\ee 
Furthermore, we take advantage of the same deformation of this expression as in (\ref{q-deform-62}), which results in 
\be 
\label{slN63}
\begin{split} 
P_r (a,q) 
& = \sum_{0\le k_3\le k_2\le k_1}^r (-1)^{k_1+k_3} {r \brack k_1}{k_1\brack k_2}{k_2\brack k_3} q^{-rk_1+\alpha(k_1,k_2,k_3)} a^{\sum_{i=1}^3 \beta_i k_i}  \times \\ 
& \qquad \qquad \times (aq^{-1};q)_{k_2} (aq^{k_3-1};q)_{k_1-k_2} (aq^{r};q)_{k_1}.
\end{split}
\ee
As usual, we fix $\beta_i$ and a quadratic polynomial $\alpha(k_1,k_2,k_3)$ by comparing with $t=-1$ specialization of the uncolored superpolynomial in table \ref{tab-P}, which can also be written as (it is again instructive to identify a zig-zag and diamonds from fig. \ref{fig-63} in this expression)
\be
P_\square (a,q,t) = 1 + a^{-1}(q^{-1}t^{-3}+t^{-2} + qt^{-1})(1+aq^{-1}t) (1+aqt^3) ,
\ee
and the $S^2$-colored polynomial
\be
\begin{split} 
P_2  (a,q) =& \,
1 - a^{-1} (1+q)(q^{-2} + 1 - q^{-1}) (1-aq^{-1})(1-aq^2) 
\\ & 
+ a^{-2}q^{-2} (1-aq^{-1}) (1-a) (1-aq^2) (1-aq^3)  
\\ & 
+ a^{-2}(q^{-5} - (1+q)q^{-4} + q^{-2}) (1-aq^{-1}) (1-a) (1-aq^2) (1-aq^3)
\\ & 
  + a^{-2} q^{-5} (1+q)(1-aq^{-1}) (1-aq^{-1}) (1-aq^2) (1-aq^3) 
\\ & 
 -a^{-2} q^{-4} (1+q)(1-aq^{-1}) (1-a) (1-aq^2) (1-aq^3).
\end{split} 
\ee
Similarly as for $6_2$ knot, the terms in the last two lines above fix the structure of $q$-Pochhammers in the second line of (\ref{slN63}), whose more complicated structure is a consequence of a horizontal displacement of diamonds in the homological diagram. From this comparison we get the final result
\be 
\begin{split} 
P_r (a,q) & = \sum_{0\le k_3\le k_2 \le k_1}^\infty 
{r\brack k_1}{k_1\brack k_2}{k_2\brack k_3}
(aq^{-1};q)_{k_2}(aq^{k_3-1};q)_{k_1-k_2} (aq^{r};q)_{k_1}  \times   \\ 
& \qquad \qquad \times (-1)^{k_1+k_3} a^{-k_1} q^{\frac12 k_1^2+\frac12 k_3^2-k_1k_2-rk_1+\frac32 k_1-k_2+\frac12 k_3} .
\end{split}
\ee 
Furthermore, an analogous computation reveals the form of colored superpolynomials
\be 
\begin{split} 
P_r (a,q,t) & = \sum_{0\le k_3\le k_2 \le k_1}^\infty
{r\brack k_1}{k_1\brack k_2}{k_2\brack k_3}
(-aq^{-1}t;q)_{k_2}(-aq^{k_3-1}t;q)_{k_1-k_2} (-aq^{r}t^3;q)_{k_1}  \times  \\ 
& \qquad \qquad \times  t^{k_3-2k_2-k_1} a^{-k_1} q^{\frac12 k_1^2+\frac12 k_3^2-k_1k_2-rk_1+\frac32 k_1-k_2+\frac12 k_3}.
\end{split}
\ee 
These results are of cyclotomic form, and are consistent with the expression in \cite{Nawata:2015wya}, which in addition proves that this expression has correct Melvin-Morton-Rozansky limit. 


\subsection{$7_4$ knot}

\begin{figure}[htp]
    \centering
    \includegraphics[width=0.3\textwidth]{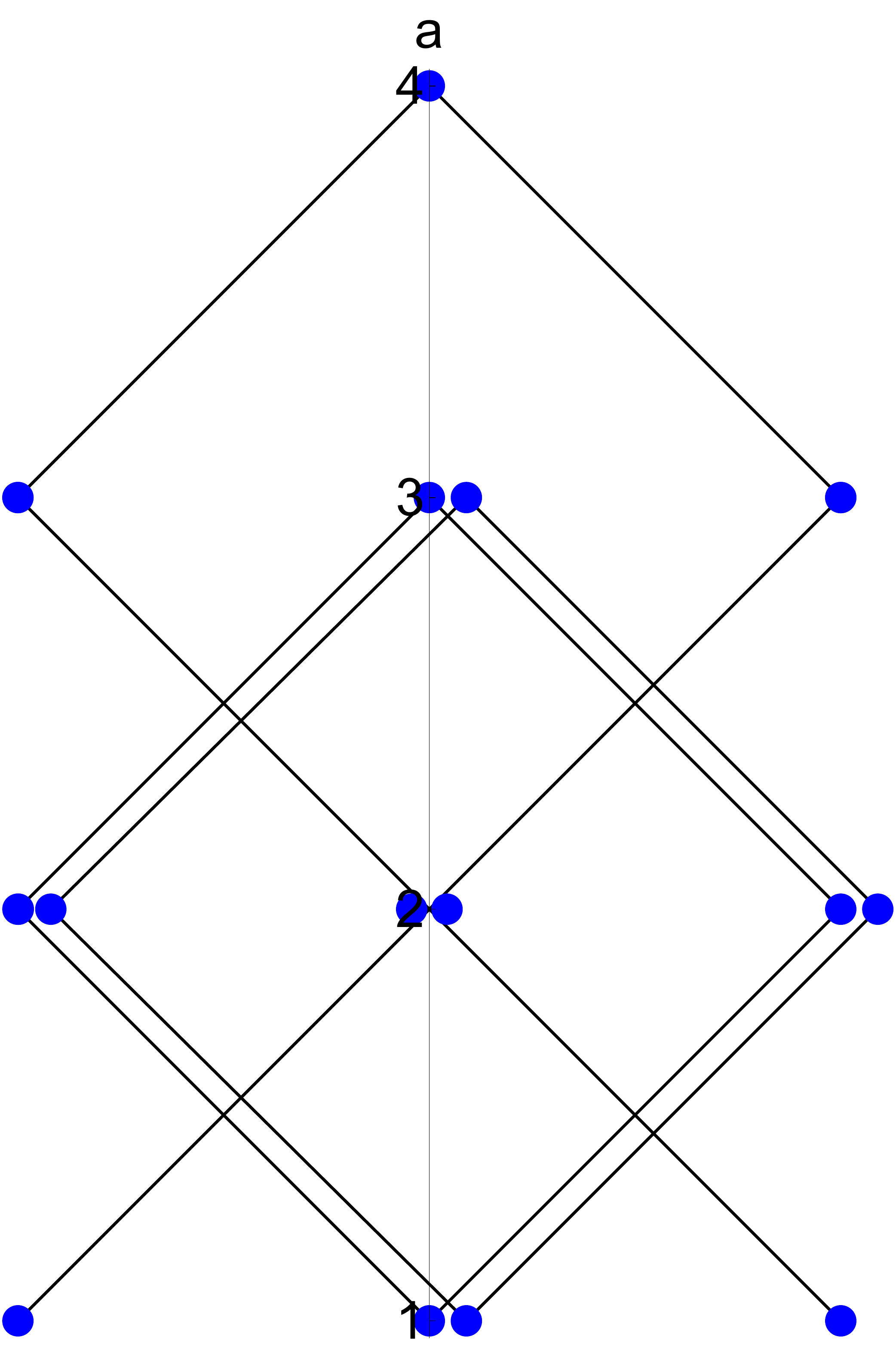} $\qquad \qquad$
    \includegraphics[width=0.3\textwidth]{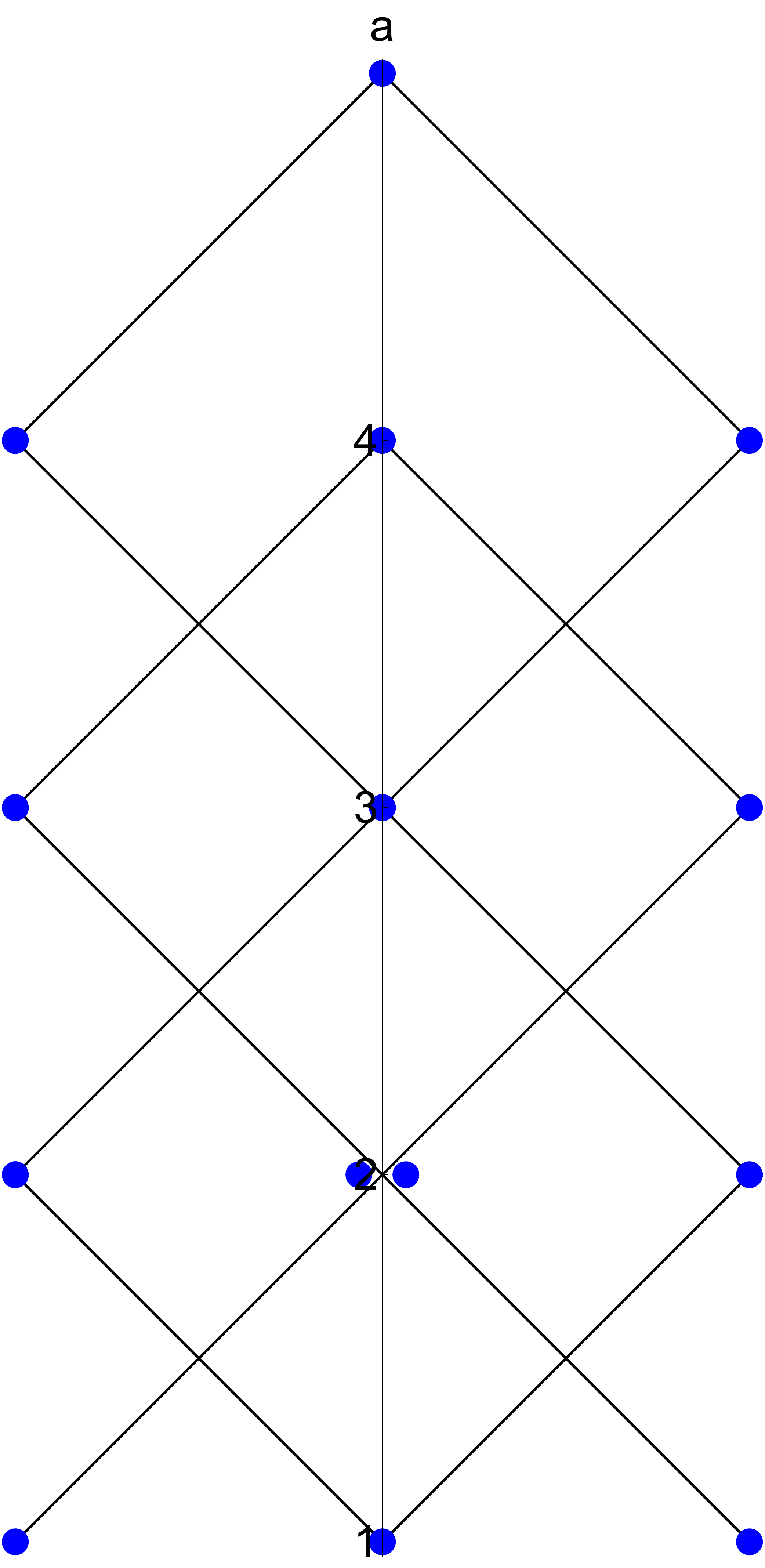}
    \caption{Homological diagram for $7_4$ (left) and $9_2$ (right) knots.}\label{fig-74}
\end{figure}

 Now we consider $7_4$ knot. Its colored HOMFLY-PT polynomials or superpolynomials have not been written explicitly before, so we may take advantage of our reconstruction scheme to provide such new results. This example is also instructive, because Alexander polynomial for $7_4$ knot is the same as for $9_2$. Homological diagrams for these two knots consist of a zig-zag of length 3 and three vertically displaced diamonds. For $9_2$ these diamonds are displaced uniformly, while for $7_4$ two of these diamonds overlap, as shown in fig. \ref{fig-74}. This difference vanishes for $a=1$, which is of course the reason why Alexander polynomials or these two knots are the same, and in the ``cyclotomic'' form they read
\be
\Delta (x) = 4x^{-1} - 7 + 4x = 1 + 4\frac{(1-x)^2}{x}.
\ee

The inverse binomial expansion then yields
\be 
\frac{1}{\Delta^{N-1}(x)} 
= \sum_{0 \le k_4 \le k_3 \le k_2 \le k_1}^\infty {N+k_1-2 \choose k_1}{k_1 \choose k_2}{k_2 \choose k_3}{k_3 \choose k_4}  (-1)^{k_1} \frac{(1-x)^{2k_1}}{x^{k_1}}.
\ee 
Taking advantage of (\ref{deform-cyclo}), our deformation procedure leads to the following form of superpolynomials
\be 
\begin{split} 
P_r(a,q) 
& = \sum_{0 \le k_4 \le k_3 \le k_2 \le k_1}^\infty (-1)^{k_1}{r \brack k_1}{k_1 \brack k_2}{k_2 \brack k_3}{k_3 \brack k_4} q^{-k_1 r + (k_1^2+k_1)/2} q^{\alpha(k_1,\ldots,k_4)}  \times\\ 
& \qquad \qquad \times 
a^{\sum_{i=1}^4 \beta_i k_i} t^{\sum_{i=1}^4 \gamma_i k_i} (-aq^{-1}t;q)_{k_1}(-aq^{r}t^3;q)_{k_1}.
\end{split}
\ee 
Interestingly, the difference in colored superpolynomials for these two knots arises only from deformation terms $\alpha(k_1,\ldots,k_4)$, $\beta_i$ and $\gamma_i$. As usual we fix them by comparing to the uncolored superpolynomials given in table \ref{tab-P}, which can also be written as follows 
\be
\begin{split}
P_\square^{7_4}(a,q,t) &= -t^{-1}+(1+aq^{-1}t) (1+aqt^3) (1+2at+a^2t^3)  , \\
P_\square^{9_2}(a,q,t) &= -t^{-1} + (1+aq^{-1}t) (1+aqt^3) (1+ at+ a^2t^3+ a^3t^5), 
\end{split}
\ee
and to the $S^2$-colored superpolynomial, which for $7_4$ knot is derived e.g. in \cite{Zodinmawia:2011oya}. This comparison fixes $\alpha(k_1,\ldots,k_4)$, $\beta_i$ and $\gamma_i$ for both knots. Ultimately we get the following colored superpolynomials for $9_2$, in agreement with results for twist knots in section \ref{ssec-twist-II}
\be 
\begin{split}
 P_r^{9_2} (a,q,t) 
& = \sum_{0 \le k_4 \le k_3 \le k_2 \le k_1}^\infty {r \brack k_1}{k_1 \brack k_2}{k_2 \brack k_3}{k_3 \brack k_4} q^{-k_1 r + (k_1^2+k_1)/2+ k_2^2+k_3^2+k_4^2-k_2-k_3-k_4} \times \\ 
& \qquad \qquad \times (-1)^{r+k_1} a^{k_2+k_3+k_4}t^{2(k_2+k_3+k_4)}  t^{-r} (-a q^{-1}t;q)_{k_1}(-aq^{r}t^3;q)_{k_1}.   \label{Pr-92}
\end{split}
\ee
On the other hand, for $7_4$ knot we obtain the following colored superpolynomials
\be
\begin{split}
P_r^{7_4} (a,q,t) 
& = \sum_{0 \le k_4 \le k_3 \le k_2 \le k_1}^\infty {r \brack k_1}{k_1 \brack k_2}{k_2 \brack k_3}{k_3 \brack k_4} q^{-k_1 r + (k_1^2+k_1)/2+ k_2^2+k_3^2+k_4^2-k_2-k_4-k_2k_3+k_2k_4-k_3k_4} \times \\ 
& \qquad \qquad \times (-1)^{r+k_1} a^{k_2+k_4}t^{2(k_2+k_4)} t^{-r} (-aq^{-1} t;q)_{k_1}(-aq^{r}t^3;q)_{k_1}.    \label{Pr-74}
\end{split}
\ee 
This result is cyclotomic, and we verified its correctness by comparison with colored superpolynomials for $r=3,4,5$ for $7_4$ knot found in \cite{Nawata:2013qpa,vivek-rama}, by checking conditions imposed by canceling differentials (\ref{eq:Cancelling}), and the consistency with the exponential growth.

In this example it is also worth illustrating the difference between the Melvin-Morton-Rozansky limit and the limit that yields A-polynomial, which is relevant for the volume conjecture. Both limits involve $q\to 1$ and $q^r=x=const$. Note that (\ref{Pr-92}) and (\ref{Pr-74}) differ only by the terms $q^{k_3-k_2 k_3 + k_2 k_4 - k_3k_4} a^{-k_3} t^{-2k_3}$ in the summand. In the Melvin-Morton-Roznasky limit we set $t=-1$ and identify $a=q^N$, so all these terms are irrelevant when we set $\hbar = \log q\to 0$, which is why we obtain the same Alexander polynomial for these knots. On the other hand, to determine A-polynomial, we need to consider saddle points of the prepotential $\widetilde{W}$, which can be determined by approximating the summations in (\ref{Pr-92}) and (\ref{Pr-74}) by integrals and introducing continuous variables $z_i=q^{k_i}$
\be 
P_r (a,q)
 \sim \int \prod_i dz_i\,  e^{\frac{1}{\hbar} (\widetilde{W}(x,z_i) + O(\hbar))}.
\ee
The difference in the summands in (\ref{Pr-92}) and (\ref{Pr-74}) implies that superpotentials associated to $9_2$ and $7_4$ knots differ by
\be
\Delta\widetilde{W} = -\log z_2 \log z_3 + \log z_2 \log z_4 - \log z_3 \log z_4 - \log a \log z_3 -2\log t\log z_3,
\ee
and these terms affect the form of Nahm equations $y = e^{x\partial_x \widetilde{W}(x,z_i)}$ and $1= e^{z_i\partial_{z_i} \widetilde{W}(x,z_i)}$ for these two knots. Furthermore, A-polynomial $A(x,y)=0$ is determined by eliminating $z_i$ from this set of equations, so this explains why A-polynomials for $7_4$ and $9_2$ knots have different form.


\subsection{$8_{19}$ knot}  \label{ssec-819}

As the final example we consider $8_{19}$ knot, or in other words $(3,4)$ torus knot. This knot is homologically thick, i.e. the  $\delta$-grading (\ref{delta-grading}) is not the same for all generators. It is therefore instructive to demonstrate that our formalism works for such knots too. 

\begin{figure}[htp]
    \centering
    \includegraphics[width=0.7\textwidth]{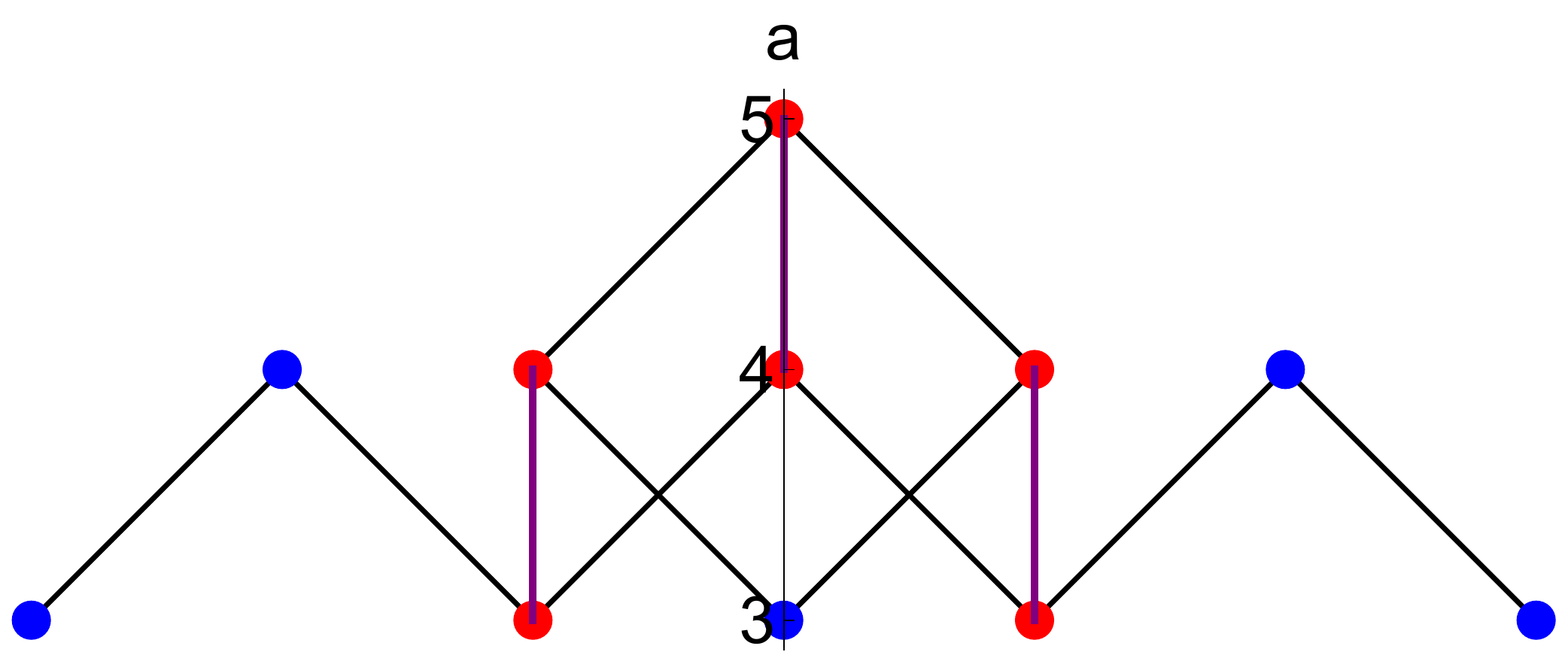}
    \caption{Homological diagram for $8_{19}$ knot. The pairs of generators shown in red, connected by vertical purple segments, correspond to monomials in the superpolynomial that are cancelled upon setting $a=1$ and $t=-1$. }\label{fig-819}
\end{figure}

For thick knots, in consequence of the action of the differential $d_0$ that reduces the HOMFLY-PT homology to the knot Floer homology, some pairs of monomials in the superpolynomial cancel upon setting $a=1$ and $t=-1$. In the homological diagram for $8_{19}$ knot shown in fig. \ref{fig-819}, the action of this differential is represented by vertical purple segments, and generators that cancel are shown in red. Therefore, even though this homological diagram consists of a zig-zag made of 7 generators and one diamond (so that the total number of HOMFLY-PT generators is 11), after canceling 3 pairs of generators, Alexander polynomial has only 5 monomial terms. Nonetheless, it can be rewritten in the form (\ref{Delta-f}) that makes the whole homological structure manifest
 \be
\begin{split}
\Delta_{8_{19}(x)} &= x^{-3}-x^{-2}+1 - x^2 + x^3 = \\ 
& 
= \underbrace{\frac{1}{x^3}\big(1-x(1-x)(1+x^2+x^4)\big)}_{\textrm{zig-zag of length 7}}  - \underbrace{\frac{1}{x}(1-x)^2}_{\textrm{diamond}} = \\ 
& 
= \frac{1}{x^3}\Big(1- x(1-x)\big(1+(1-x)x+ x^2(1+x^2)\big)\Big).
\end{split}
\ee 
Now, following our ``homological'' approach, the inverse binomial expansion yields
\be 
\begin{split}
& \frac{1}{\Delta^{N-1}_{8_{19}}(x)}   
= x^{3(N-1)}\sum_{k_1=0}^\infty {N+k_1-2 \choose k_1} x^{2k_1}(1-x)^{2k_1} \big(1+x^{-1}(1-x)^{-1}(1+x^2+x^4)\big)^{k_1} =
\\ &  \quad 
=  x^{3(N-1)}\sum_{0\le k_2\le k_1}^\infty {N+k_1-2 \choose k_1}{k_1\choose k_2} x^{2k_1-k_2}(1-x)^{2k_1-k_2} (1+x^2+x^4)^{k_2} =
\\ & \quad
=  x^{3(N-1)}\sum_{0\le k_4\le k_3\le k_2\le k_1}^\infty {N+k_1-2 \choose k_1}{k_1\choose k_2}{k_2\choose k_3}{k_3\choose k_4} x^{2k_1-k_2+2(k_3+k_4)}(1-x)^{2k_1-k_2}.
\end{split}
\ee 
Then, $q$- and $a$-deformation of this expression leads to the following form of colored HOMFLY-PT polynomial
\be
\begin{split} 
P_r (a,q) & = \frac{a^{3r}}{q^{3r}} \sum_{0\le k_4\le k_3\le k_2\le k_1}^r {r \brack k_1}{k_1\brack k_2}{k_2\brack k_3}{k_3\brack k_4} q^{r(2k_1-k_2+2(k_3+k_4))}q^{\sum_{i,j=1}^4 \beta_{ij}k_i k_j+\sum_{i=1}^4\gamma_i k_i}
\\ & 
\qquad \qquad \qquad \qquad  \times 
a^{\sum_{i=1}^4 \delta_i k_i} (a/q;q)_{k_1}(aq^r,q)_{k_1-k_2},
\end{split}
\ee 
with parameters $\beta_{ij}$, $\gamma_i$ and $\delta_i$ still to be fixed. We fix them by comparing with the uncolored HOMFLY-PT polynomial 
\be 
P_\square (a,q,t) = 
 a^3q^{-3} + (1+atq^{-1}) a^3q^{-1}t^2 (1+q^2t^2+q^4t^4) + (1+aqt^3)(1+atq^{-1} ) a^3 t^4
\ee 
(which is also given in table \ref{tab-P}) and $S^2$-colored polynomial, whose form follows from the $S^2$-colored superpolynomial
\be 
\begin{split}
P_2(a,q,t) =&\,  a^6 q^{-6} + (1+atq^{-1} ) (1+aq^2t^3) a^6q^{-2}t^2 (1+q^6 t^4+q^{12}t^8)
\\ & 
+ (1+q) (1+atq^{-1})( 1+aq^2 t^3) a^6qt^4 (1+q^2t^2+q^4t^4+q^6t^6)
\\ & 
+(1+atq^{-1}) a^7q^{-4}t^3 (1+q^3t^2+q^6t^4+q^9t^6+q^{12}t^8 + q^{15} t^{10})
\\ & 
+ (1+q)(1+atq^{-1}) (1+aq^2t^3) (1+aq^3t^3) a^6 q^{-1}t^4 (1+q^3t^2+q^6t^4) 
\\ & 
+ (1+atq^{-1}) (1+at) (1+aq^2t^3) (1+aq^3t^3) a^6q^4 t^8.
\end{split} 
\ee 
From this comparison we find 
\be
\begin{split}
& \beta_{11} = \beta_{22} = \frac12, \qquad \beta_{33}=\beta_{44} = \beta_{24} = 1,\qquad
\beta_{12}=\beta_{23}=0, \qquad \beta_{13}=\beta_{14}=\beta_{34}=-1, 
\\ & 
\gamma_1= 1, \qquad\gamma_2= 0, \qquad\gamma_3 = \frac12, \qquad \gamma_4=-\frac12,\qquad
\delta_1=\delta_2=\delta_3=\delta_4 = 0, 
\end{split}
\ee 
so that colored HOMFLY-PT polynomials for this knot take form 
\be
\begin{split} 
P_r (a,q) & = \frac{a^{3r}}{q^{3r}} \sum_{0\le k_4\le k_3\le k_2\le k_1}^r {r \brack k_1}{k_1\brack k_2}{k_2\brack k_3}{k_3\brack k_4} q^{r(2k_1-k_2+2(k_3+k_4))} \times
\\ & 
 \qquad    \times 
q^{\frac12(k_1^2+k_2^2)+k_3^2+k_4^2-k_1k_3-k_1k_4+k_2k_4-k_3k_4} q^{k_1+\frac12(k_3-k_4)} (aq^{-1};q)_{k_1}(aq^r,q)_{k_1-k_2}.
\end{split}
\ee 
We can also introduce $t$-dependence, and analogously determine colored superpolynomials 
\be
\begin{split} 
P_r (a,q,t) & = \frac{a^{3r}}{q^{3r}} \sum_{0\le k_4\le k_3\le k_2\le k_1}^r {r \brack k_1}{k_1\brack k_2}{k_2\brack k_3}{k_3\brack k_4} t^{k_1+k_2+2(k_3+k_4)} q^{r(2k_1-k_2+2(k_3+k_4))} \times
\\ & 
 \quad  \times 
q^{\frac12(k_1^2+k_2^2)+k_3^2+k_4^2-k_1k_3-k_1k_4+k_2k_4-k_3k_4} q^{k_1+\frac12(k_3-k_4)} (-atq^{-1};q)_{k_1}(-aq^rt^3,q)_{k_1-k_2}.
\end{split}   \label{Praqt-819}
\ee 
This expression matches the result in \cite{Gukov:2015gmm}. Following our second approach we could also derive the cyclotomic form of colored superpolynomials, which is given in \cite{Gukov:2015gmm} too. We have also confirmed validity of (\ref{Praqt-819}) from other perspectives discussed earlier; in particular we verified that it satisfies the relation (\ref{eq:Cancelling}) with Rasmussen invariant $s=-3$.


\section*{Acknowledgements}
We thank for discussions and acknowledge insightful comments from Tobias Ekholm, Stavros Garoufalidis, Angus Gruen, Sergei Gukov, Piotr Kucharski, Satoshi Nawata, Sunghyuk Park, Ramadevi Pichai, Vivek Singh and Marko Sto\v si\' c. S.B. is supported by Humboldt postdoctoral grant. He also acknowledges his stay at the University of Warsaw where this project was initiated. J.J.  was  supported by the Polish National Science Centre (NCN) grant 2016/23/D/ST2/03125.  The work of P.S. is supported by the TEAM programme of the Foundation for Polish Science co-financed by the European Union under the European Regional Development Fund (POIR.04.04.00-00-5C55/17-00).

\newpage
\bibliographystyle{JHEP}
\bibliography{bibdynam}

\providecommand{\href}[2]{#2}\begingroup\raggedright\begin{thebibliography}{10}

\bibitem{Melvin}
P.~Melvin and H.~Morton, {\it {The coloured Jones function}},  {\em Comm. Math.
  Phys., 169 (1995), no. 3, 501–520}.

\bibitem{Rozansky}
L.~Rozansky, {\it {A contribution of the trivial connection to the Jones
  polynomial and Witten’s invariant of 3d manifolds}},  {\em I, Comm. Math.
  Phys., 175 (1996), no. 2, 275–296}.

\bibitem{Rozansky1}
L.~Rozansky, {\it {Higher order terms in the Melvin-Morton expansion of the
  colored Jones polynomial}},  {\em Commun. Math. Phys.} {\bf 183} (1997),
  no.~2 291--306.

\bibitem{Rozansky2}
L.~Rozansky, {\it {The universal R-matrix, Burau representation, and the
  Melvin-Morton expansion of the colored Jones polynomial}},  {\em Adv. Math.,
  134 (1998), no. 1, 1–31}.

\bibitem{Bar-Natan1996}
D.~Bar-Natan and S.~Garoufalidis, {\it {On the Melvin-Morton-Rozansky
  conjecture}},  {\em Inventiones mathematicae} {\bf 125} (1996), no.~1
  103--133.

\bibitem{garoufalidis2005analytic}
S.~Garoufalidis and T.~T.~Q. Le, {\it An analytic version of the
  melvin-morton-rozansky conjecture},
  \href{http://arxiv.org/abs/math/0503641}{{\tt math/0503641}}.

\bibitem{Dunfield:2005si}
N.~M. Dunfield, S.~Gukov, and J.~Rasmussen, {\it {The superpolynomial for knot
  homologies}},  \href{http://arxiv.org/abs/math/0505662}{{\tt math/0505662}}.

\bibitem{Gukov:2011ry}
S.~Gukov and M.~Stosic, {\it {Homological Algebra of Knots and BPS States}},
  {\em Proc. Symp. Pure Math.} {\bf 85} (2012) 125--172,
  [\href{http://arxiv.org/abs/1112.0030}{{\tt arXiv:1112.0030}}]. [Geom. Topol.
  Monographs18,309(2012)].

\bibitem{Fuji:2012pm}
H.~Awata, S.~Gukov, P.~Sulkowski, and H.~Fuji, {\it {Volume Conjecture: Refined
  and Categorified}},  {\em Adv. Theor. Math. Phys.} {\bf 16} (2012), no.~6
  1669--1777, [\href{http://arxiv.org/abs/1203.2182}{{\tt arXiv:1203.2182}}].

\bibitem{Fuji:2012pi}
H.~Fuji, S.~Gukov, M.~Stosic, and P.~Sulkowski, {\it {3d analogs of
  Argyres-Douglas theories and knot homologies}},  {\em JHEP} {\bf 01} (2013)
  175, [\href{http://arxiv.org/abs/1209.1416}{{\tt arXiv:1209.1416}}].

\bibitem{Nawata:2012pg}
S.~Nawata, P.~Ramadevi, Zodinmawia, and X.~Sun, {\it {Super-A-polynomials for
  Twist Knots}},  {\em JHEP} {\bf 11} (2012) 157,
  [\href{http://arxiv.org/abs/1209.1409}{{\tt arXiv:1209.1409}}].

\bibitem{Nawata:2015wya}
S.~Nawata and A.~Oblomkov, {\it {Lectures on knot homology}},  {\em Contemp.
  Math.} {\bf 680} (2016) 137, [\href{http://arxiv.org/abs/1510.01795}{{\tt
  arXiv:1510.01795}}].

\bibitem{Gukov:2015gmm}
S.~Gukov, S.~Nawata, I.~Saberi, M.~Stosic, and P.~Sulkowski, {\it {Sequencing
  BPS Spectra}},  {\em JHEP} {\bf 03} (2016) 004,
  [\href{http://arxiv.org/abs/1512.07883}{{\tt arXiv:1512.07883}}].

\bibitem{Gukov:2019mnk}
S.~Gukov and C.~Manolescu, {\it {A two-variable series for knot complements}},
  \href{http://arxiv.org/abs/1904.06057}{{\tt arXiv:1904.06057}}.

\bibitem{Park:2019xey}
S.~Park, {\it {Higher rank $\hat{Z}$ and $F_K$}},
  \href{http://arxiv.org/abs/1909.13002}{{\tt arXiv:1909.13002}}.

\bibitem{Ekholm:2020lqy}
T.~Ekholm, A.~Gruen, S.~Gukov, P.~Kucharski, S.~Park, and P.~Sulkowski, {\it
  {$\widehat{Z}$ at large $N$: from curve counts to quantum modularity}},
  \href{http://arxiv.org/abs/2005.13349}{{\tt arXiv:2005.13349}}.

\bibitem{Habiro_2007}
K.~Habiro, {\it {A unified Witten-Reshetikhin-Turaev invariant for integral
  homology spheres}},  {\em Inventiones Mathematicae} {\bf 171} (2007), no.~1
  1–81.

\bibitem{berest2019cyclotomic}
Y.~Berest, J.~Gallagher, and P.~Samuelson, {\it {Cyclotomic Expansion of
  Generalized Jones Polynomials}},  \href{http://arxiv.org/abs/1908.04415}{{\tt
  arXiv:1908.04415}}.

\bibitem{Kucharski:2017poe}
P.~Kucharski, M.~Reineke, M.~Stosic, and P.~Sulkowski, {\it {BPS states, knots
  and quivers}},  {\em Phys. Rev.} {\bf D96} (2017), no.~12 121902,
  [\href{http://arxiv.org/abs/1707.02991}{{\tt arXiv:1707.02991}}].

\bibitem{Kucharski:2017ogk}
P.~Kucharski, M.~Reineke, M.~Stosic, and P.~Sulkowski, {\it {Knots-quivers
  correspondence}},  {\em Adv. Theor. Math. Phys.} {\bf 23} (2019) 1849--1902,
  [\href{http://arxiv.org/abs/1707.04017}{{\tt arXiv:1707.04017}}].

\bibitem{Fuji:2012nx}
H.~Fuji, S.~Gukov, and P.~Sulkowski, {\it {Super-A-polynomial for knots and BPS
  states}},  {\em Nucl. Phys.} {\bf B867} (2013) 506--546,
  [\href{http://arxiv.org/abs/1205.1515}{{\tt arXiv:1205.1515}}].

\bibitem{Fuji:2013rra}
H.~Fuji and P.~Sulkowski, {\it {Super-A-polynomial}},  {\em Proc. Symp. Pure
  Math.} {\bf 90} (2015) 277--304, [\href{http://arxiv.org/abs/1303.3709}{{\tt
  arXiv:1303.3709}}].

\bibitem{Aganagic:2012jb}
M.~Aganagic and C.~Vafa, {\it {Large N Duality, Mirror Symmetry, and a
  Q-deformed A-polynomial for Knots}},
  \href{http://arxiv.org/abs/1204.4709}{{\tt arXiv:1204.4709}}.

\bibitem{manolescu2014introduction}
C.~Manolescu, {\it {An introduction to knot Floer homology}},
  \href{http://arxiv.org/abs/1401.7107}{{\tt arXiv:1401.7107}}.

\bibitem{Gorsky:2013jxa}
E.~Gorsky, S.~Gukov, and M.~Stosic, {\it {Quadruply-graded colored homology of
  knots}},  \href{http://arxiv.org/abs/1304.3481}{{\tt arXiv:1304.3481}}.

\bibitem{Kontsevich:2010px}
M.~Kontsevich and Y.~Soibelman, {\it {Cohomological Hall algebra, exponential
  Hodge structures and motivic Donaldson-Thomas invariants}},  {\em Commun.
  Num. Theor. Phys.} {\bf 5} (2011) 231--352,
  [\href{http://arxiv.org/abs/1006.2706}{{\tt arXiv:1006.2706}}].

\bibitem{reineke2011degenerate}
M.~Reineke, {\it {Degenerate Cohomological Hall algebra and quantized
  Donaldson-Thomas invariants for m-loop quivers}},
  \href{http://arxiv.org/abs/1102.3978}{{\tt arXiv:1102.3978}}.

\bibitem{Panfil:2018sis}
M.~Panfil, M.~Stosic, and P.~Sulkowski, {\it {Donaldson-Thomas invariants,
  torus knots, and lattice paths}},  {\em Phys. Rev. D} {\bf 98} (2018), no.~2
  026022, [\href{http://arxiv.org/abs/1802.04573}{{\tt arXiv:1802.04573}}].

\bibitem{Stosic:2017wno}
M.~Stosic and P.~Wedrich, {\it {Rational links and DT invariants of quivers}},
  \href{http://arxiv.org/abs/1711.03333}{{\tt arXiv:1711.03333}}.

\bibitem{Stosic:2020xwn}
M.~Stosic and P.~Wedrich, {\it {Tangle addition and the knots-quivers
  correspondence}},  \href{http://arxiv.org/abs/2004.10837}{{\tt
  arXiv:2004.10837}}.

\bibitem{Ekholm:2018eee}
T.~Ekholm, P.~Kucharski, and P.~Longhi, {\it {Physics and geometry of
  knots-quivers correspondence}},  \href{http://arxiv.org/abs/1811.03110}{{\tt
  arXiv:1811.03110}}.

\bibitem{Ekholm:2019lmb}
T.~Ekholm, P.~Kucharski, and P.~Longhi, {\it {Multi-cover skeins, quivers, and
  3d $\mathcal{N}=2$ dualities}},  {\em JHEP} {\bf 02} (2020) 018,
  [\href{http://arxiv.org/abs/1910.06193}{{\tt arXiv:1910.06193}}].

\bibitem{Panfil:2018faz}
M.~Panfil and P.~Sulkowski, {\it {Topological strings, strips and quivers}},
  {\em JHEP} {\bf 01} (2019) 124, [\href{http://arxiv.org/abs/1811.03556}{{\tt
  arXiv:1811.03556}}].

\bibitem{Zodinmawia:2011oya}
Zodinmawia and P.~Ramadevi, {\it {SU(N) quantum Racah coefficients \& non-torus
  links}},  {\em Nucl. Phys. B} {\bf 870} (2013) 205--242,
  [\href{http://arxiv.org/abs/1107.3918}{{\tt arXiv:1107.3918}}].

\bibitem{Nawata:2013qpa}
S.~Nawata, P.~Ramadevi, and Zodinmawia, {\it {Colored HOMFLY polynomials from
  Chern-Simons theory}},  {\em J. Knot Theor.} {\bf 22} (2013) 1350078,
  [\href{http://arxiv.org/abs/1302.5144}{{\tt arXiv:1302.5144}}].

\bibitem{vivek-rama}
P.~Ramadevi and V.~Singh, {\it 2020},  {\em {Private communication}}.

\end{thebibliography}\endgroup

\end{document}